\documentclass[11pt]{article}

\usepackage{epsfig,amsfonts,amssymb,amscd,amsmath,mathrsfs,xspace,mathrsfs,amsthm,dsfont,bbm}
\usepackage[small,bf,hang]{caption}
\usepackage{mathabx,amsmath}  
\usepackage{slashed}
\usepackage{cite} 
\usepackage{esint}  

\usepackage{comment}
\usepackage{color}

\usepackage{bbold}

\usepackage{hyperref}

\newcommand{\sectiono}[1]{\section{#1}\setcounter{equation}{0}}

\newcommand{\be}{\begin{equation}}
\newcommand{\ee}{\end{equation}}
\newcommand{\ben}{\begin{eqnarray}\displaystyle}
\newcommand{\een}{\end{eqnarray}}
\newcommand{\bea}{\begin{eqnarray}}
\newcommand{\eea}{\end{eqnarray}}

\newcommand{\refb}[1]{(\ref{#1})}

\def\ZZZ{{\hbox{ Z\kern-1.6mm Z}}}
\def\RRR{{\hbox{ R\kern-2.4mm R}}}
\def\CCC{{\hbox{ C\kern-2.0mm C}}}
\def\zzz{{\hbox{z\kern-1mm z}}}

\def\ZZZ{{\mathbb Z} }
\def\RRR{{\mathbb R} }
\def\CCC{{\mathbb C} }
\def\cc{{\bf b}}  
\def\AL{{\eta_c}}

\newcommand{\Qr}{{q_r}}

\newcommand{\p}{\partial}

\def\({\left(}
\def\){\right)}
\def\[{\left[}
\def\]{\right]}

\newcommand{\non}{\nonumber}

\newcommand{\qeq}{{\hbox{=\kern-2.3mm ? \kern.5mm }}}
\renewcommand{\qeq}{=}

\newcommand{\VV}{{\cal V}}
\newcommand{\BB}{{\cal B}}

\newcommand{\BBB}{{\cal B}}

\newcommand{\AAA}{{\cal A}}

\newcommand{\MM}{{\cal M}}

\newcommand{\wh}{\widehat}

\newcommand{\TT}{{\cal T}}

\def\cl0{\tilde c_0}

\newcommand{\Hom}{\widehat\Omega}

\def\one{{\hbox{ 1\kern-.8mm l}}}
\def\zero{{\hbox{ 0\kern-1.5mm 0}}}

\newcommand{\bra}[1]{\ensuremath{\langle {#1} |}}
\newcommand{\ket}[1]{\ensuremath{| {#1} \rangle}}

\newcommand{\inc}{{\cdots}}

\begin{document}

\begin{titlepage}   
	\rightline{}
	\rightline\today 
	\rightline{MIT-CTP/5701} 
	\begin{center}
		\vskip 1.1cm

  {\Large 
		 \bf{On the normalization of open-closed string amplitudes} } \\[1.0ex]
		\vskip 1.5cm
		
{\large\bf {Ashoke Sen$^\dag$ and Barton Zwiebach$^*$}}
		\vskip 1cm
		
		$^\dag$ {\it   International Centre for  Theoretical Sciences -- TIFR,\\
			Bengaluru - 560089, India}\\
		
		\vskip .3cm
		
		$^*$ {\it   Center for Theoretical Physics, \\
		Massachusetts Institute of Technology, \\
		Cambridge MA 02139, USA}\\
		\vskip .1cm
		
		\vskip .4cm
		ashoke.sen@icts.res.in, zwiebach@mit.edu

		\vskip 1.9cm
		\end{center}
	
\begin{quote} 	
		
\centerline{\bf Abstract} 

\medskip

We use the factorization constraints of open-closed string field theory to 
determine the signs and normalizations 
of general string amplitudes with both open
and closed string  external states. 
The normalization of all amplitudes is controlled by the genus, the number of
boundaries, the number of open and closed string
insertions, the string coupling and the D-brane tension. 
The challenge with signs arises because the relevant 
moduli spaces are not complex manifolds 
and have no obvious orientation. 
We deal with this by fixing a specific convention for the 
sign of the integration measure over the moduli space and 
adopting a consistent prescription for the ordering of operators and ghost 
insertions inside correlators. 
	\end{quote} 
	\vfill
	\setcounter{footnote}{0}
	
	\setcounter{tocdepth}{2}  
	
\end{titlepage}

\baselineskip 15pt 

\tableofcontents

\sectiono{Introduction}

Open-closed superstring amplitudes
are important in string theory model building, 
since the backgrounds often contain D-branes and orientifolds 
of various types.  
Even in the absence of background D-branes, 
open-closed amplitudes are needed for computing D-instanton corrections to
purely closed string amplitudes.
A special set of open-closed amplitudes 
includes tree-level pure open string amplitudes as well 
as all the amplitudes of the pure quantum closed string  
theory.  
The normalization and sign factors for those special
amplitudes are known, although not usually discussed in
detail.   As we will explain here, the normalization
and sign factors of  {\em general} open-closed amplitudes
do not seem to be known, except
in some special
cases\cite{Polchinski:1998rq,Polchinski:1998rr,Balthazar:2019rnh,Sen:2021tpp,
Eniceicu:2022xvk,Alexandrov:2023fvb}.  
Finding this missing information
is the goal of this paper. While this  
is a question in string perturbation theory,  we will 
arrive at the answer using string field theory.

After the formulation of bosonic tree-level open 
string theory\cite{Witten:1985cc} and that of quantum
closed string theory~\cite{Zwiebach:1992ie}, 
bosonic open-closed string field theory was 
formulated~\cite{Zwiebach:1990qj,Zwiebach:1992bw,Zwiebach:1997fe}
as a regular version of quantum open string theory, a theory with
both open and closed string fields.  The construction was extended to 
open-closed superstring theory 
in~\cite{FarooghMoosavian:2019yke}.\footnote{Some recent applications
of open closed string field theory can be found in 
\cite{Maccaferri:2022yzy,Maccaferri:2023gcg,Maccaferri:2023gof,Firat:2023gfn}.}
In both bosonic and superstring versions of open-closed string field theory,
a key ingredient in the construction are forms on the relevant moduli spaces of
Riemann surfaces.  When integrated over the moduli spaces, they give 
amplitudes.  
The lack of known normalization factors for open-closed forms implies
open-closed string field theory is missing details in its definition, resulting  
in a missing `prescription' for the world-sheet rules for perturbative
open-closed amplitudes.  
In fact, for the issue of normalization
we can work with on-shell amplitudes, and all vertex operators can
be taken to be on-shell.   We focus on bosonic closed and open-closed
theories, because the notation is simpler.  The normalization factors for closed
bosonic strings apply for heterotic
and type II strings, and the normalization factors for open-closed
bosonic theory apply to open-closed superstrings.

This paper is organized as follows. In section~\ref{sconv} 
we first describe the conventions that we shall use for writing
string amplitudes
and then 
describe our results for the
normalization
of these amplitudes.
In section \ref{backgroundinfo} we review some of the basic features of string field theory
and how the Feynman diagrams of string field theory generate world-sheet results for
string amplitudes. In section \ref{NormForm} 
we use this construction to find a set of recursion relations
for the
normalization constants appearing in different string amplitudes. In section
\ref{sanalysis} we solve these recursion relations to determine the normalization constants 
explicitly. In section \ref{samplitudes} we use our results to write down explicit 
expressions for some simple string amplitudes. In appendix \ref{sa} we 
derive a result in closed string field theory that helps
relate the constant that appears 
in the normalization of correlation functions on the upper half plane 
to the D-brane tension.
In appendix~\ref{sb} we describe the 
computation of the D-brane tension in bosonic
string theory in flat space-time using the results of this paper.
We shall use $\alpha'=1$ units throughout this paper. 

\sectiono{Conventions and summary of results} \label{sconv}

In this section we shall describe our normalization conventions for the
correlation functions in the world-sheet theory and give expressions
for string amplitudes in terms of these correlation functions up to some
overall numerical constants. Our goal in the later sections will be to
determine these numerical constants.

\subsection{String amplitudes}

We begin by discussing how
Riemann surfaces
are built by gluing.  For Riemann surfaces $\Sigma_{g,b,n_c,n_o}$ of genus
$g$,  $b$ boundary components, $n_c$ closed string punctures, and $n_o$ open 
string punctures, the surface can be decomposed into the following regions: 
\begin{enumerate}

\item $n_c$ disjoint disks $D_1, \ldots, D_{n_c} $, one around each closed
string puncture.  The boundaries of the disks are called gluing circles.

\item $n_o$ disjoint half-disks $d_1, \ldots, d_{n_o} $, one around each open
string puncture.  The puncture and the diameter of the half disk lie on the
boundary.  The half-circle border of each disk is called 
a gluing segment.

\item A number $\# S$ of   $2g-2+n$ spheres 
$S_1, \ldots, S_{\# S}$,
each with three holes or gluing circles.

\item A number $\# H$ of `hexagonal' disks $H_1, \cdots , H_{\#H}$  
with three gluing segments and three segments 
that are part of the boundary, 
-- henceforth referred to as boundary segments. 
The gluing and boundary segments 
alternate as we travel around the hexagon.\footnote{We can view 
the hexagon as the result of slicing a sphere with three holes into half in a way that also
cuts each hole into two intervals, and identifying the hexagon with one of the halves. 
In this case
the boundaries that were part of the holes are the gluing segments and the boundaries
that arise as the result of slicing are the boundary segments.}

\item A number $\# A$ of annuli  $A_1, \cdots, A_{\# A}$ with one hole being a gluing circle and
the other hole part of the boundary of the surface.    

\item A number $\#\bar A$ 
of annuli $\bar A_1, \cdots, \bar A_{\# \bar A}$ with one hole being a gluing circle and
the other hole being composed of a boundary segment and a gluing segment.

\end{enumerate}

The various components listed above are given complex coordinates. 

\begin{enumerate}
\item  There are local coordinates $w_a$ on each disk
$D_a$, chosen so that the closed string puncture is at  $w_a=0$.   
There are local coordinates $\eta_i$ on each half disk $d_i$, 
so that the  open string puncture is at $\eta_i=0$, and the locus
$\{ \hbox{Im} (\eta_i) =0, \, -1 \leq \hbox{Re} (\eta_i)\leq 1\}$ is
on the boundary of the surface.  

\item  There are coordinates $z_i$ on the sphere $S_i$,
coordinates $\rho_a$ and $\bar \rho_a$ on the annuli $A_a$ 
and $\bar A_a$, respectively, 
and coordinates $\xi_i$ on each hexagonal disk $H_i$

\end{enumerate}

\begin{figure}[h]
	\centering
\epsfysize=5.6cm
\epsfbox{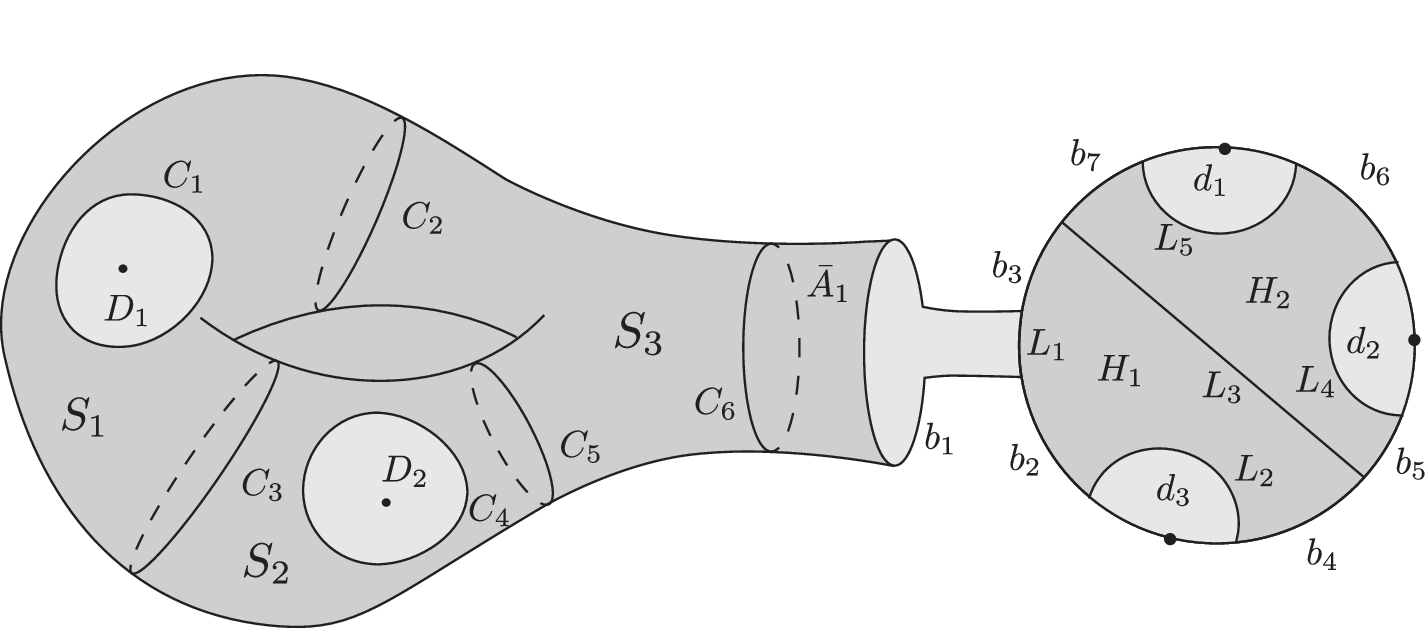}
	\caption{\small 
	A surface of $g=1, b=1, n_c=2, n_o=3$ built 
	with two disks $D_1, D_2$, 
	three half-disks $d_1, d_2, d_3$, 
	three spheres $S_1, S_2, S_3$, one barred annulus $\bar A_1$, and two hexagonal disks
	$H_1, H_2$. They are joined across 
	six gluing circles $C_1,\cdots ,C_6$
	and five gluing segments $L_1,\cdots L_5$.
	The $b_i$'s as well as the segments of $d_i$'s containing the open string punctures
	make up the single boundary.} 
	\label{FF1FF}
\end{figure}

The gluing that makes up the surface is done across gluing circles and gluing segments
using transition functions for the relevant coordinates: 

\begin{enumerate}

\item We have $\# C$ gluing circles
$C_1, \ldots, C_{\# C}$.  The gluing can join disk
to sphere,  disk to annulus,  sphere to sphere, sphere to annulus, or
annulus to annulus across their gluing circles. 
The annuli here can be of either type. 
The gluing can also join two gluing circles 
of the same sphere.
For each gluing circle there is a  
transition function that relates the coordinates on two sides of the gluing circle.

\item
We have $\# L$ gluing segments 
$L_1, \ldots, L_{\# L}$.  The gluing can join half-disks to hexagons,
or half-disks the segment in $\bar A$ annuli.  Hexagons can glue
to hexagons and to segments in $\bar A$ annuli.  
Furthermore, gluing can
join two gluing segments of the same hexagon.
For each gluing segment there is a  
transition function that relates the coordinates on two sides of the segment.  
\end{enumerate}  
For illustration, we show in Figure~\ref{FF1FF} a genus one surface with one
boundary component, constructed using all the types of regions introduced
above, except for annuli of $A$ type. 

The moduli of the Riemann surface are described 
by these transition functions 
modulo
non-singular redefinitions of the coordinates (while keeping the closed string
punctures at $w_a=0$ and open string punctures at $\eta_i=0$).
 This data is sufficient to define on-shell  
amplitudes, but off-shell amplitudes are not invariant under reparametrization of
$w_a$ and the $\eta_i$ and so we need the information on the choice of $w_a$ 
(up to its phase) and $\eta_i$.
The decomposition of the Riemann surface  
into disks, spheres, annuli, half-disks, and hexagons, described above is not unique, but 
answers for physical quantities
do not depend on what decomposition we use.  
A given surface may
be built with different number of ingredient regions, the
annulus with one open string puncture, for example, can be built 
with either
$\{ A_1, \bar A_1, d_1\}$ or with $\{ H_1, d_1\}$.

\medskip 
\noindent

The expressions for string amplitudes require antighost insertions
$\BB[\p/\p u^i]$ for any tangent vector $\p/\p u^i$ in the moduli space, 
so we now review their
definition.  
Let $\{C_s\}$ and $\{L_m\}$  
denote,  respectively, all the gluing circles and the
gluing segments
that are involved in building the Riemann surface from its components.
Let us choose some particular orientations along the gluing circles $C_s$ and
the gluing segments $L_m$. Furthermore, 
let $\sigma_s$ and $\tau_s$ be, respectively,
complex coordinates on the left and right 
of the oriented gluing circle~$C_s$ and 
$\sigma_m$ and $\tau_m$ be, respectively,
complex coordinates on the left and right 
of the oriented gluing segment $L_m$. Then the transition functions across
the gluing circles and gluing segments take the form:
\be
\label{eo765u} 
\sigma_s=F_s(\tau_s, u)\, ,
\ee
and 
\be
\label{open-str-gluing} 
\sigma_m = G_m(\tau_m, u)\,,
\ee
where $u$ stands for the collection of all the moduli.
Associated with a modulus $u^i$, we 
introduce the following linear combination of the anti-ghost fields $b,\bar b$:
\be \label{e233open}
\begin{split}  
\BB\left[{\p\over \p u^i}\right] \equiv \ & \   \sum_s \ \biggl[ \ \ointop_{C_s} {\p F_s\over \p u^i}
d\sigma_s b(\sigma_s) + \ointop_{C_s} {\p \bar F_s\over \p u^i}
d\bar\sigma_s \bar b(\bar \sigma_s) \biggr]    
\\
&\hskip-10pt   +
\sum_m \ \biggl[ \ \intop_{L_m}   
{\p G_m\over \p u^i}
d\sigma_m b(\sigma_m) + \intop_{L_m}   
{\p \bar G_m\over \p u^i}
d\bar\sigma_m \bar b(\bar \sigma_m) \ 
\biggr]\,. 
\end{split}
\ee 
In the integrals we use the normalization
$\ointop_0 {dz\over z} = \ointop_0 {d\bar z\over \bar z} = 1$.
This completes the definition of the `antighost insertion' $\BB$. 

\medskip
 The amplitudes in  open-closed string theory are computed
for a given genus $g$ and a given number $b$ of boundary components. 
Moreover, we have $n_c$ closed string vertex operators 
$A^c_1,\cdots, A^c_{n_c}$ and $n_o$ open 
string vertex operators $A^o_1,\cdots, A^o_{n_o}$. We shall take
all of these to be unintegrated vertex operators described by 
dimension zero primaries.  The amplitudes are given by
\ben 
\label{e549new} 
&& \hskip-30pt
\AAA_{g,b,n_c,n_o}(A^c_1,\cdots, A^c_{n_c};A^o_1,\cdots, A^o_{n_o}) \nonumber \\
&=& \  
 g_s^{-\chi_{g,b,n_c,n_o}} \, N_{g,b,n_c,n_o}\, 
  \int_{\MM_{g,b,n_c,n_o}}    
\hskip-20pt \Hom^{(g,b,n_c,n_o)} 
(A^c_1,\cdots, A^c_{n_c}; A^o_1,\cdots, A^o_{n_o})  \, ,
\een
where $g_s$ is the string coupling, $\chi_{g,b,n_c,n_o}$ is the Euler number of the surfaces,
\be
\label{o-c-euler}
\chi_{g,b,n_c,n_o} =  2 - 2g -n_c - b - \tfrac{1}{2} n_o\,,
\ee
$N_{g,b,n_c,n_o}$ is a normalization constant,
and the $\Hom^{(g,b,n_c,n_o)}$ are top forms in the moduli space,
thus of degree equal to the real dimension of the moduli space
\be 
\label{dimrealMGN}
\hbox{dim}\, \MM_{g,b, n_c, n_o} = 6g-6+3b+2n_c+n_o
\equiv p\,. 
\ee
We call these forms canonical forms, and denote them with a hat.  They are     
given by
\be\label{edefOmegaOpen}
\begin{split}
&\Hom^{(g,b,n_c,n_o)} (A_1^c,\cdots, A_{n_c}^c; A_1^o, \cdots , A_{n_o}^o ) \\
& \qquad \equiv \left\langle \BB
\left[{\p\over \p u^{j_1}}\right] \cdots
\BB\left[{\p\over \p u^{j_p}}\right]  \, A_1^c \cdots A_{n_c}^c \ 
A_1^o \cdots A_{n_o}^o  
\right \rangle_{{}_{\Sigma_{g,b,n_c,n_o} }}  
\hskip-25pt du^{j_1} \wedge \cdots \wedge du^{j_p} \, .
\end{split} 
\ee
The genus $g$ surface $\Sigma_{g,b,n_c,n_o }$ has $n_c$ closed string punctures 
distributed in the bulk and $b$ boundary components,
with the $n_o$ open string punctures distributed among them.
$\MM_{g,b,n_c,n_o}$ is the moduli space of such surfaces and $\{u^i\}$ are the
real coordinates on this moduli space.
The closed string vertex operators $A^c_1,\cdots, A^c_{n_c}$ are inserted at the
closed string punctures and the open string vertex operators $A^o_1,\cdots, A^o_{n_o}$
are inserted at the open string punctures.
The determination of the normalization constants 
$N_{g,b,n_c,n_o}$
is the main goal of this paper.
We shall do this by
using the gluing relations between amplitudes that are the basis
of string field theory.  

An exception to the definition~\refb{edefOmegaOpen} of canonical
forms
will be made for the disk one point
function of a closed string. In this case the surface has
a conformal Killing vector and has no moduli.
It is therefore convenient to introduce the zero form
\be \label{e14}
\Hom^{(0,1,1,0)}(A^c) \equiv -{1\over 2\pi i} \,  
\langle c_0^-A^c\rangle\, ,
\ee
where we define 
\be
c_0^\pm\equiv \tfrac{1}{2} \, (c_0\pm \bar c_0), \qquad
b_0^\pm\equiv  (b_0\pm \bar b_0), \qquad L_0^\pm\equiv  (L_0\pm \bar L_0)\, .
\ee
With no moduli to integrate over and with Euler number zero,
the associated amplitude~is 
\be \label{eckv}
\AAA_{0,1,1,0}(A^c)= N_{0,1,1,0} \, \Hom^{(0,1,1,0)}(A^c)\, .
\ee

A general off shell string state is described by vertex operators that are not
conformally invariant.
The prescription given above for computing on-shell amplitudes can be generalized
to define off-shell amplitudes if we specify the coordinate system in which we insert
the off-shell vertex operators. We take these to be the coordinates $w_a$ on the
disks surrounding the closed string punctures or the half-disks containing the open
string punctures.

\subsection{Integrated vertex operators}

We can use the construction described above to see how integrated vertex
operators arise from \refb{e549new}. 
First
let us consider the case of an on-shell open string
vertex operator $cV^o$. 
Let $w$ be a local
coordinate at the puncture $w=0$ where the open string is inserted, and let
$z$  
be the coordinate system in a
patch of the surface that shares a gluing segment with the 
half-disk $d$ 
that contains the puncture at $w=0$. Let
\be
\label{forppp}
z=G(w,u)\, ,
\ee
be the transition function across this gluing segment. We shall assume that $G(w,u)$ is
analytic in the half-disk~$d$. 
Then $y=G(0,u)$ is the location of the
puncture in the $z$ coordinate, with $u$ playing the role of a modulus. 
We shall assume that all other transition functions 
are independent of $u$ so that $u$ can be regarded as the modulus 
associated with the
location of the open string puncture at $z=y$.
The effect of inserting the vertex operator $cV^o$ is
represented by taking the wedge product with
the canonical one-form 
\be
\label{homform1} 
\Hom_1  (c V^o ) 
=    \, du \,  \BB
\left[{\p\over \p u}\right]   \,    c V^o (w=0)  
  \, ,
\ee
inside the correlation function in \refb{edefOmegaOpen}.
Note that we have taken $\BB\left[{\p\over \p u}\right] $
to be positioned immediately to the left of the
vertex operator, but there may be additional signs that arise from moving the $\BB$'s from
the arrangement given in \refb{edefOmegaOpen} to the configuration given above.

We now use the doubling trick to express the sum of the holomorphic and
 anti-holomorphic integrals along open contours in $\BB$,
 as given in the second line of \refb{e233open},
 into integration over a closed contour surrounding the 
 puncture at $w=0$ or equivalently at $z= y$.  
 Note that~\refb{forppp} is of the form of \refb{open-str-gluing}.
 Since $cV^o$ is a dimension zero primary we have $cV^o(w=0) = cV^o(z=y)$ 
 and therefore we have   
\be \label{ebboneopen}  
\Hom_1 (c V^o )  
 = \, du \, \ointclockwise b(z) dz {\p G(w,u)\over \p u}\  c V^o (z=y)  \, .
 \ee 
Given the prescription that the 
$w$ coordinate system must be to the right of the contour,
 the contour is oriented clockwise around the position of the puncture.
 This is indicated by the $\ointclockwise$ symbol. 
The integral then gives 
 \be \label{ebboneopenxx}
 \Hom_1 (cV^o) 
 = \  -  \, du \  {\partial y\over\partial u}  V^o (z=y) =  - \,  dy \,  V^o(y)\,, 
  \ee
with  
$V^o(y)$ 
understood as the vertex operator at $z=y$ using the local
coordinate $(z-y)$ vanishing at that insertion point.

A similar analysis can be done for on-shell closed string vertex operators of the form
$c\bar c V^c$. The result is a two-form that is given by 
\be\label{e243}
\Hom_2 (c \bar c V^c ) = \ - dy\wedge d\bar y \, V^c(y)\, ,   
\ee
with the minus sign arising from having to rearrange  $\BB[\p/\p u]\BB[\p/\p\bar u]
c\bar c V^c$ as  $-\BB[\p/\p u] c \BB[\p/\p\bar u] \bar c V^c$ and then applying the
manipulations given above.

\subsection{Normalization and sign conventions}\label{norm-conve}

In order to state our results for $N_{g,b,n_c,n_o}$,
we need to describe our conventions for various
quantities that go into the definition of the amplitudes 
and the forms  
$\Hom^{(g,b,n_c,n_o)}_{p}$. We shall
describe this below.
\begin{enumerate}
\item In calculating the closed string amplitudes on the complex plane, we shall 
use the convention
\ben
\label{sl2ccorr}
&& \hskip-30pt \langle  c(z_1) \bar c (\bar z_1)c (z_2) \bar c(\bar z_2) c(z_3) \bar c(\bar z_3) e^{ik\cdot X}(z) 
\rangle \nonumber \\[1.0ex]
&=& \hskip-5pt - (2\pi)^D \delta^{(D)} ( k)\, (z_1-z_2) (z_2-z_3) (z_1-z_3) 
(\bar z _1-\bar z _2) (\bar z _2-\bar z _3) (\bar z _1-\bar z _3) \,. 
\een
Here $D$ is the number of non-compact dimensions.
We are allowed to choose the normalization this way because we can rescale the
ghost operators $c(z)$ and $\bar c(\bar z)$ by a constant, 
while doing a compensating rescaling
of the antighosts $b(z)$ and $\bar b(\bar z)$ to preserve the  
$b\,c$ operator product expansion.  In this
way, any additional multiplicative 
constant on the right hand side of the above overlap can be eliminated by 
rescaling the ghosts.  \refb{sl2ccorr} can also be written as
\be
\label{sl2c_overlap}
\langle k | c_{-1} \bar c_{-1} c_0 \bar c_0 c_1 \bar c_1|k'\rangle 
= - (2\pi)^D \delta^{(D)} ( k + k')\,. 
\ee

\item Next we consider the normalization of the correlation functions on the
upper half plane. For this we shall assume that the background under study
contains a single D$p$-brane whose world-volume extends along $(p+1)$ non-compact
directions and possibly some compact directions whose details will not be
needed for our analysis.\footnote{We shall denote by D$p$-brane 
any D-brane whose
world-volume extends along $p+1$ non-compact space-time coordinates,
with Dirichlet boundary conditions along any additional non-compact direction. 
The boundary conditions along the compact directions can depend on the description,
{\it e.g.} T-duality along a circle switches Dirichlet to Neumann boundary condition
and vice versa. For a general CFT associated with the compact directions there
may not be a clear distinction between Dirichlet, Neumann and mixed boundary
conditions. We regard all of these as different kinds of D$p$-branes.}
We take,
\be \label{eopencorr}
\langle c(z_1) c(z_2) c(z_3) e^{ik.X}(0) \rangle = - K  \, 
(2\pi)^{p+1}
\delta^{(p+1)}(k)  \ (z_1-z_2) (z_2-z_3) (z_1-z_3)\, .  
\ee
 The value of the (possibly complex) constant $K$ that appears on the right-hand side
must be determined in the theory.  Indeed, having
used the scaling of ghosts and antighosts to 
set the normalization of the sphere amplitude as in \refb{sl2ccorr},
we are no longer able to
change the value of $K$ above.   Clearly, the value of $K$ affects all disk
amplitudes; they are all proportional to $K$. 
Since surfaces with boundaries can be built by 
gluing disks, the constant $K$ also affects all amplitudes with boundaries.  

The need for having the constant $K$ can be seen as follows.
Let us consider   
the graviton one-point function on the 
disk in the presence of a D$p$-brane.  
The graviton vertex operator will be
insensitive to  the boundary condition 
along the compact directions since it only involves the scalar fields
associated with the
non-compact coordinates.  
Therefore the disk one-point function of such
a graviton will be independent of the boundary condition along the compact
directions except for the factor of $K$. 
On the other hand 
this one-point function is expected to be 
proportional to the D-brane tension which does 
depend on the boundary condition
along the compact direction,    
{\it e.g.} our D$p$ brane with Neumann or 
Dirichlet boundary conditions along  an internal circle 
are different D$p$ branes, have different tensions, and hence should 
give different result for this one point function.  
Such a dependence comes through  $K$.\footnote{In principle 
we could take $K=1$ and absorb
the normalization constants in $N_{g,b,n_c,n_o}$. However, we shall find it
more convenient to work in a convention in which $N_{g,b,n_c,n_o}$ is
universal,  independent of the background.}  
The one-point function being proportional to the D-brane tension 
thus implies that $K$ is proportional to the D-brane tension. 
We shall describe this relation in \refb{eKTrel}.

Equation~\refb{eopencorr} 
gives the following matrix element between the  two momentum states built on
the SL$(2, \mathbb{R})$ invariant open string
vacua:
\be\label{sl2r_overlap}
\langle k| c_{-1} c_0 c_1|k'\rangle 
= -(2\pi)^{p+1} \ K  
\delta^{(p+1)}(k+k')\, .
\ee

Once we have fixed the normalization of the sphere and the disk correlation functions, this in
principle determines the normalization of the correlation
functions on all other Riemann surfaces
since we can insert complete set of closed  (open) string states across the gluing
circles (gluing segments) and express them in terms of sphere and disk amplitudes.
The consistency of conformal field theory ensures that the result is independent of the
representation of the Riemann surface that we use\cite{segal-g,Sonoda:1988mf,
Sonoda:1988fq}.   
The results can be readily generalized to the cases
of multiple D-branes.

\item It follows from the definition of amplitudes in~\refb{e549new}
that we can change $N_{g,b,n_c,n_o}$ by redefining $g_s$ by a multiplicative constant.
We shall use this freedom to set $N_{0,0,3,0}=1$, so that the closed string three-point
function on the sphere is given by $g_s$ times 
the three-point function of the corresponding vertex 
operators, without any additional normalization factor.

\item For
target-space fields that are real,   
we take the associated
physical state vertex operators of 
closed and open strings to be of the form 
\be \label{edefacao}
A^c =c\bar c V^c, \qquad A^o=cV^o\, ,
\ee
where $V^c(z,k)$ and $V^o(x,k)$ are, respectively,
 dimension (1,1) and dimension 1 matter primaries carrying momentum $k$,
normalized by the following operator products 
\be \label{evertexnorm}
V^c(z_1, -k) V^c(z_2, k) \simeq {4\over |z_1-z_2|^4}, \qquad V^o(x_1,-k) V^o(x_2, k)\simeq
{K^{-1}\over (x_1-x_2)^2}\, .  
\ee
The reason for choosing the normalizations this way will be explained in 
section \ref{skinetic}. 
We shall often make use of the open string vertex operator $W^o$ 
with $K$ independent norm
\be \label{edefW}
W^o\equiv K^{1/2} V^o, \qquad W^o(x_1, -k) W^o(x_2, k) 
\simeq {1\over (x_1-x_2)^2}, 
\qquad A^o = K^{-1/2} \, c\, W^o\, . 
\ee

\item Next we need to fix the sign factor for the measure of integration.   
In order to get a number after integrating a form
 over the moduli space of punctured Riemann surfaces, we need to specify
an orientation for the moduli space; alternatively, we just have
to state if the integral of the form $du^1\wedge \cdots\wedge
du^p$ gives a positive result or a negative result. 
Happily, this orientation can be fixed naturally if
we have only external closed
strings and no boundaries 
since the moduli space of Riemann surfaces without boundaries
has a natural complex
structure.  For complex moduli 
$u=u_x+iu_y$ we can take $du\wedge d\bar u=-2i du_x\wedge d u_y$,
with integration over $du_x\wedge du_y$ giving 
positive result.  This applies also to the moduli 
associated to the position $z = x + i y$ of a puncture; here $dx\wedge dy$ is
declared to give a positive integral.  

Once we consider amplitudes with  boundaries, the situation
becomes more complicated. 
The moduli spaces of Riemann surfaces in open-closed theory
do not have a complex structure. In fact such moduli
 spaces can be odd-dimensional.
For a moving open string puncture, for example, we have one real modulus, 
for a boundary we generally have three real moduli. 
In addition, 
open string vertex operators are Grassmann odd, and changing
their positions relative to each other and relative to the $\BB$ insertions
will also lead to a change
of the sign of the above correlator.  Unless we fix definite conventions for the
signs of the integration measure and the  positioning of the operators inside the
correlation function, 
the normalization
constants $N_{g,b,n_c,n_o}$ will have ambiguous signs. Below we shall describe
the specific conventions that we use to resolve these ambiguities.
\end{enumerate}

In order to fix the sign conventions in open - closed string amplitudes,
we follow
{\em three prescriptions}  for the insertions
and correlators.
The prescriptions begin with the known forms on surfaces
without boundaries and with only closed string insertions.  Boundaries
and open string vertex operators are recursively inserted as follows:
\begin{enumerate}

\item[{P1.}]  {\em Adding boundaries.}  To add a boundary to 
$\wh\Omega^{(g,b,n_c,n_o)}$   
we begin with 
a form $\wh\Omega^{(g,b,n_c+1, n_o)}$ with one extra  
closed string puncture, assumed to be known.  
We then
define the one-form closed string state $\ket{\cc}$ 
given by  
\be\label{ebinsertX}
\ket{\cc}  \equiv
 \,  -\, {d\Qr\over \Qr}\, (e^{-\Lambda}\Qr)^{L_0^+} b_0^+
|B\rangle\, ,   \ \ \   \Qr \in [0, 1]\,, 
\ee
where  $\Lambda$ is some positive
constant and $\ket{B}$ is the ghost number three boundary state, defined so that
for any closed string state $|\chi\rangle$, $\langle B|\chi\rangle$ denotes the disk one
point function of $\chi$, with $\chi$ inserted in a coordinate system in which 
the disk has unit radius. 
The orientation  of the integration measure over $\Qr$ 
is  along the direction of increasing $\Qr$.
We then define the desired form with $b+1$ boundaries by inserting $\ket{\cc}$
on the added puncture: 
\be\label{eomegadef} 
\Hom^{(g,b+1,n_c,n_o)} ( \cdots )  \equiv   
\Hom^{(g,b,n_c+1, n_o)} (\cdots,  \ket{\cc} ) \,, 
\ee
where the dots represent all other
operators inserted elsewhere. 
Since $|B\rangle$ has ghost number three, $b_0^+|B\rangle$ is a Grassmann even state
and hence we can insert it anywhere inside the correlation function.
The reason for the above definition of $\ket{\cc}$ will
become apparent as we later establish~\refb{e433}.

Let $|\Sigma\rangle$ be the closed string state such that 
for any closed string state $\ket{\chi}$ we have
\be \label{e428a}  
 \Hom^{(g,b,n_c+1, n_o)} (\cdots,  \ket{\chi} ) =\langle\Sigma|\chi\rangle\, .
\ee
Then, the form in~\refb{eomegadef} can be expressed as
\be\label{esewtoA} 
\Hom^{(g,b+1,n_c,n_o)} ( \cdots )  =   - \, 
\Big\langle\Sigma\Big|  {d\Qr\over \Qr}\, (e^{-\Lambda}\Qr)^{L_0^+} b_0^+ 
\Big|B\Big\rangle
= \Big\langle B\Big|  {d\Qr\over \Qr}\, (e^{-\Lambda}\Qr)^{L_0^+} b_0^+ \Big|\Sigma
\Big\rangle  \, .
\ee
The extra minus sign in the second step 
comes from having to move $b_0^+$ through the 
Grassmann odd operator $B$ in using the BPZ exchange
property for the overlap.
Recalling
that $\langle B|\chi\rangle$ is the disk one point function of $\chi$
for any closed string state $|\chi\rangle$, the
above expression is the disk one point function 
of the state ${d\Qr\over \Qr}\, (e^{-\Lambda}\Qr)^{L_0+\bar L_0} 
b_0^+ \Big|\Sigma\rangle$.

The form constructed in~\refb{esewtoA}   
can be given the following
geometric interpretation (see Figure~\ref{FF2FF}).  
Let $w_1$ denote the local coordinate at the $(n_c+1)$-th 
 puncture of the original Riemann surface
 $\Sigma$  
to be sewn to the one puncture disk, and $w_2$ denote the coordinate on the disk in which the disk has unit radius. We now
identify the  
coordinates $w_1$ and  $w_2$  via 
\be\label{ebsew}
w_1w_2 = e^{-\Lambda} \, \Qr\, .
\ee 
Let $\Sigma_B$
denote the resulting Riemann surface.\footnote{
Since the boundary of the disk was at $|w_2|=1$, in the $w_1$ coordinate 
the boundary
 is at
$|w_1|=e^{-\Lambda}\Qr$. Therefore
in the $w_1$ coordinate system $\Sigma_B$ appears as
$\Sigma$ with a hole of size  $e^{-\Lambda}|q_r|$ around $w_1=0$.} 
Then  we can interpret \refb{esewtoA} as a correlator on~$\Sigma_B$
\be
\label{sigme9kd} 
\Hom^{(g,b+1,n_c,n_o)} ( \cdots )  =
{d\Qr\over \Qr}\, \langle b_0^+ (\cdots) \rangle_{\Sigma_B}\, ,
\ee
where $\cdots$ denotes the product of
vertex operators and $b$-insertions present on the original Riemann surface and
$b_0^+$ integration contour is placed 
next to the boundary at $|w_2|=1$, 
oriented 
such that it keeps the boundary to the right. 
It is to be
understood that the sign of \refb{sigme9kd} is determined from 
\refb{esewtoA}, which is unambiguous.
This is important since, as written, \refb{sigme9kd} suffers from 
apparent sign  ambiguities due to the presence of unintegrated
Grassmann odd open string vertex operators and possibly
unpaired $\BB$ insertions hidden in the $(\cdots)$.
\begin{figure}[h]
	\centering
\epsfysize=7.7cm
\epsfbox{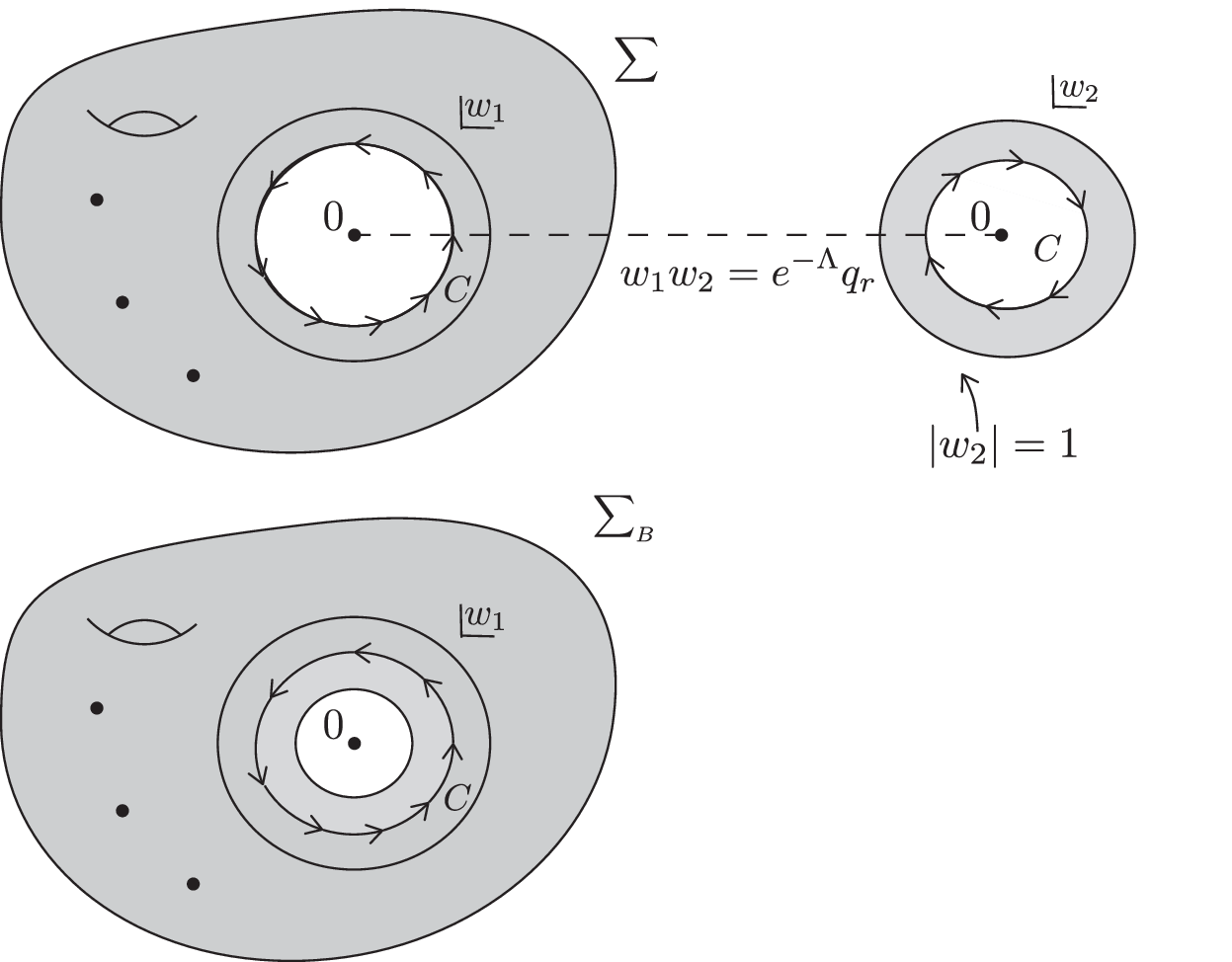}
	\caption{\small   
	The puncture $w_1=0$ on the surface $\Sigma$ is used to glue
	a disk $|w_2| \leq 1$ with a puncture at $w_2=0$.  This gluing,
	with $w_1 w_2 = e^{-\Lambda} q_r$ generates a boundary in
	$\Sigma$ giving us the surface $\Sigma_B$, shown below. 
	The gluing circle $C$ 
	is also used 
	to support the integration~\refb{e224n} for the antighost
	insertion.} 
	\label{FF2FF}
\end{figure}
\medskip 
\noindent

We can relate $b_0^+$ to the expected $\BB[\p/\p \Qr]$ insertion as follows.
In \refb{ebsew}, identifying $w_2$ as $\sigma_s$ and $w_1$ as $\tau_s$ 
in the language used for the insertion~\refb{e233open}, we have 
$w_2 = G ( w_1 , \Qr)$,  with $G ( w_1 , \Qr)=e^{-\Lambda} \Qr/w_1$. This gives
${\partial G\over \partial q_r} = e^{-\Lambda}/w_1=w_2/q_r$, 
and hence  
\be \label{e224n}
\BB\left[{\p\over \p \Qr}\right] = \ointclockwise dw_2 \, b(w_2) \,  {w_2\over \Qr} 
+\ointclockwise  d\bar w_2 \, \bar b(\bar w_2) \, {\bar w_2\over \Qr}=-{b_0^+\over \Qr}
\, ,
\ee
where in the first step we have used that the integration 
contour $C$ must keep the region covered by the $w_2$ coordinate system to the
left. Using this,  we 
express~\refb{sigme9kd} 
as,  
\be\label{esewtoB}
\Hom^{(g,b+1,n_c,n_o)} ( \cdots )  = -\, 
\Big\langle {d\Qr}\, 
\BB\left[{\p\over \p \Qr}\right] (\cdots) 
\Big\rangle_{\Sigma_B}\, .
\ee
The minus sign on the right hand side of \refb{esewtoB} means 
that prescription P1
implies 
the following prescription 
for the forms :
{\em insert $\BB[\p/\p \Qr]$ into the correlator  
with $\Qr$ integrated
in the direction of decreasing $\Qr$.}  
Alternatively, we could  insert $(-\BB[\p/\p \Qr])$  
in the correlation function and
integrate $\Qr$ along the
direction of increasing $\Qr$.
As long as the other operators in the correlator are even -- 
either closed string vertex operators or $\BB$-$\bar\BB$ combinations, it does not
matter where we place $\BB[\p/\p \Qr]$ inside the correlation function. 
However, if there are odd operators present in the correlator then 
we need to follow the prescription \refb{esewtoA} with $b_0^+/q_r$ replaced by
$\BB[\p/\p \Qr]$ or $(-\BB[\p/\p \Qr])$ depending on the orientation 
of the $\Qr$ integration.
The integration contour that 
defines $\BB[\p/\p \Qr]$ is placed next to
the boundary, 
without any other operator sitting between the integration contour and the
boundary.  

When the gluing parameter $\Qr$ decreases,
 the vertex operators inserted on the original Riemann surface and
carrying fixed $w_1$ coordinates, approach each other 
in the $w_2$ coordinate system. 
This picture is useful since
we can now replace  $\Qr$ by any other coordinate $y$ in \refb{esewtoB} and as long
as the increasing $\Qr$ corresponds to increasing $y$, the formula will remain valid.
Otherwise we get an extra minus sign. This observation will allow us to fix the sign
in any coordinate system on the moduli space
without having to transform the expression to the coordinate system $\Qr$.

\item[{P2.}]  {\em Adding integrated open string operators.} 
 To insert in an amplitude an on-shell open string vertex operator $cV^o$ 
we insert in the correlator the operator 
\be
\label{integopeninsert}
dy \, V^o(y) \,,
\ee
with $y$ the location of the operator on a boundary and integration along
$y$ is oriented such that as 
$y$ increases, 
the operator moves along the boundary keeping the surface to the left.  
The meaning of $V^o(y)$ in \refb{integopeninsert} is provided by 
a slight generalization of the discussion below~\refb{ebboneopenxx}. 
Let $z$ denote the complex coordinate of the Riemann surface whose restriction
to the boundary is the coordinate $y$. 
Even though $V^o$ is only defined on the boundary, 
via analytic continuation we can extend this to a holomorphic operator $V^o(z)$ defined
in the neighborhood of $y$, inserted using the coordinate system $z$. 
$V^o(y)\, dy$ is then taken to be the restriction of
the  conformally invariant holomorphic one form $V^o(z) \, dz$ 
to the
boundary.\footnote{The necessity of regarding $V^o(y)$ as the restriction of a holomorphic
operator can be seen through the following example.
Let us work in the upper half $z$ plane and let $y$ denote the real part of $z$.
Consider now a coordinate transformation $z'= -z$, $y'= -y$ that takes the upper
half $z$ plane to the lower half $z'$ plane.
Under this coordinate transformation we have $\int_{-\infty}^\infty V^o(y) dy =
-\int_{-\infty}^\infty V^o(y') dy'$ {\it after taking into account the fact that $V^o$, regarded
as a holomorphic dimension one operator,
picks up a minus sign when we go from
the $z$ coordinate to the $z'$ coordinate}.
This agrees with \refb{integopeninsert}, since in the lower half $z'$ plane
integration along the real axis keeps the surface to the right.
In contrast, if we had regarded $V^o$ as the restriction of a dimension (1/2,1/2)
operator in interpreting its conformal transformation property, 
then there would be no minus sign under $z'=-z$ transformation.
}

The prescription \refb{integopeninsert}, for convenience, 
does not include the sign in the 
one-form~\refb{ebboneopenxx}, because 
as noted below~\refb{homform1}, 
this sign is itself 
dependent on the position of $\BB$ inside the correlator. 
It follows from \refb{ebboneopenxx} that the prescription of inserting
\refb{integopeninsert}
corresponds to  the following prescription for   
the forms: 
{\em  integrate $u$ along the boundary contour  
that keeps the surface to the left and
insert $(-\BB[\p/\p u])$ 
immediately to the
left of the vertex operator $c V^o(y)$, or equivalently, insert 
$\BB[\p/\p u]$  immediately to the
right of the vertex operator $c V^o(y)$.}     
This prescription holds for any external open string state, 
not necessarily on-shell. Since $V_0(y)$ (or equivalently the combination
of $\BB[\p/\p u]$ and the unintegrated vertex operator) is Grassmann even, there is no
ambiguity arising from where we place the operator inside the correlation function and we can
use this prescription for any number of external open strings.

\item[{P3.}] {\em Exceptions.} \ 
 Prescriptions P1 and P2 
 have exceptions if there are unintegrated open string
vertex operators
or boundary sizes are not integrated.  
This problem is specific to $g=0$ with 
$n_c+b< 3$; for $n_c+b\ge 3$ 
we can fix the locations of  
three of the closed string vertex operators and/or boundaries, and integrate over
{\em all} the open string positions and boundary sizes.
 
There are three relevant cases.  In the first two cases the disk boundary sizes 
are not integrated.  All three cases feature unintegrated open string vertex
operators.  
\begin{itemize}
\item  Disk amplitudes with $n_o\geq 3$. 
Here the three unintegrated vertex operators are placed
inside the correlator in the  order they appear as we travel along the boundary keeping
the Riemann surface to the left. The rest of the open string operators are inserted
following the prescription P2.  

\item The disk with one open string and 
one closed string. 
We prescribe that $\Hom$ 
comes with an extra
sign, i.e.\ we use 
\be\label{esignchange}  \
\Hom^{(0,1,1,1)} = - \, \langle c V^o\, c\bar c V^c\rangle\, .
\ee
Any further open string insertion is treated using P2.

It is instructive to compare \refb{esignchange} with what we would get if we had
applied \refb{integopeninsert} on the closed string one point function on the disk to
compute the amplitude. Using 
\refb{e14} we see that the expected result is:
\be\label{ealtsign}
\Hom^{(0,1,1,1)} =
-{1\over 2\pi i} \ \int_{-\infty}^\infty dy\, \langle V^o(y) \, c_0^- c\bar c V^c(i) \rangle\, ,
\ee 
where the correlator is to be evaluated in the upper half plane. Using
$\rm SL(2,\mathbb{R})$ invariance
we can use the transformation $z \to z' = (z-y)/(1+ zy)$ to 
move the vertex operator from $y$ to $0$.  This replaces 
$V^o(y)$ by $V^o(0)/(1+y^2)$ in
the correlation function, and there is no effect on $c_0^- c\bar c V^c(i) $ because
this operator is conformal invariant and $z=i$ maps to $z'=i$. 
The $y$ integral now gives $\pi$ and we have
\be\label{ealtsignX}
\Hom^{(0,1,1,1)} =
{i\over 2 } \ \, \langle V^o(0) \, c_0^- c\bar c V^c(i) \rangle\, ,
\ee 
The ghost part of the correlator here can be computed using~\refb{eopencorr} noting
that the doubling prescription instructs one to replace $\bar c (z) \to c(z^*)$.  We find
\be\label{ealtsignXX}
\Hom^{(0,1,1,1)} =
2iK \ \, \langle V^o(0)  V^c(i) \rangle\, ,
\ee 
where the matter correlator is normalized with $\langle e^{ik.X}\rangle = (2\pi)^{p+1} 
\delta^{(p+1)} (k)$.
 We can now consider the postulated~\refb{esignchange}.  Fixing the vertex operators, both of which are conformal invariant, at $0$ and $i$ we have
\be\label{esignchangeX}  \
\Hom^{(0,1,1,1)} = - \, \langle c V^o (0)\, c\bar c V^c (i)\rangle =
2iK \ \, \langle V^o(0)  V^c(i) \rangle\, ,
\ee
after computing the ghost correlator.  The agreement
with~\refb{ealtsignXX} provides a justification for the minus
sign in \refb{esignchange}.
We shall later provide more confirmation 
that this minus 
sign is needed for compatibility with~\refb{integopeninsert}.

\item 
For an annulus 
with one external open string, we 
follow P1, and build $\Hom$ 
by  inserting $\ket{\cc}$
at the closed string puncture in the
open-closed disk amplitude \refb{esignchange}. 
Again,
any further open string insertion can be treated using  P2.
\end{itemize}  

\end{enumerate} 

\subsection{Main results} 

In these conventions, our result is that
the coefficients $N_{g,b,n_c,n_o}$ are given by
\be
\label{esolrecnewNNX}
N_{g, b\, n_c,  n_o} =  \AL^{3g-3+n_c + {3\over 2} b + {3\over 4} n_o}\, ,
\qquad 
\AL \equiv -{1\over 2\pi i}\, .
\ee 
Back in~\refb{e549new} this fixes the amplitudes of the open-closed theory, with
the canonical forms $\Hom$ written following the prescriptions P1, P2, and P3. 
The factor $\AL$ has its origin in the nature of the closed-string propagator.

The above values of the $N$ coefficients 
are also part of the calculation that shows
that the constant $K$ normalizing upper-half plane correlators 
is related  to the string coupling $g_s$ and the brane tension $\TT$
as follows
\be\label{eKTrel}
{K} = -g_s    {\TT\over 2\sqrt \AL}\, .
\ee
Moreover, the open string coupling $g_o$ 
can be expressed in terms of $K$ and $g_s$:
\be\label{edefgoint}
g_o^2 =  g_s \,  {K}^{-1} \, \AL^{3/2}\, .
\ee
The last two equations
 imply the familiar relation~\cite{Sen:1999xm}:
\be\label{e3100int}
 \TT\, g_o^2 =   {1\over 2\pi^2} \,.
\ee

It is also useful to state the result 
for open-closed amplitudes
using normalized forms, denoted with a prime, and defined as follows
\be \label{edefHomp}
\Hom^{'(g,b,n_c,n_o)}
(A^c_1,\cdots A^c_{n_c}; A^o_1,\cdots, A^o_{n_o}) 
\equiv K^{-b}\, \Hom^{(g,b,n_c,n_o)}(A^c_1,\cdots A^c_{n_c}; A^o_1,\cdots, A^o_{n_o})\, .
\ee
Using the relations $A^c_i=c\bar c V_i$, $A^o_i
=K^{-1/2} c W^o_i$
and 
\refb{esolrecnewNNX}-\refb{e3100int}, we can express
the amplitudes in~\refb{e549new}, \refb{eckv} as,  
\ben  
\label{e549newer} 
&& \hskip -30pt \AAA_{g,b,n_c,n_o}(A^c_1,\cdots, A^c_{n_c};A^o_1,\cdots, A^o_{n_o}) \nonumber \\[0.8ex]
&\hskip -50pt =& \hskip -30pt \  
 g_s^{2g-2+2b+n_c} \, \left( -\tfrac{\TT}{2}\right)^b \, g_o^{n_o}\, \AL^{3g-3+b+n_c}\, 
  \int_{\MM_{g,b,n_c,n_o}}    
\hskip-20pt \Hom^{\prime (g,b,n_c,n_o)} 
(c\bar c V^c_1,\cdots, c\bar c V^c_{n_c}; c W^o_1,\cdots, c W^o_{n_o})  \, .
\een

An alternative version of the above result  is a rearrangement that 
emphasizes the role of surfaces of type $(g,b,n_c, 0)$ in controlling
the prefactor.  We have 
\be  
\label{e549newerXX} 
\begin{split}
  \hskip-10pt\AAA_{g,b,n_c,n_o}(A^c_1,\cdots, A^c_{n_c};A^o_1,\cdots, A^o_{n_o}) 
 = \,  \  &  
 g_s^{-\chi_{g,b,n_c,0}} \ g_o^{n_o} K^b \  \AL^{{1\over 2}
  {\rm dim}_R \MM_{g,b,n_c, 0}}
     \\[0.5ex]
  & \hskip-10pt \cdot\int_{\MM_{g,b,n_c,n_o}} \hskip-30pt\Hom^{\prime (g,b,n_c,n_o)} 
(c\bar c V^c_1,\cdots, c\bar c V^c_{n_c}; c W^o_1,\cdots, c W^o_{n_o})  \, .
\end{split} 
\ee

The following three sections in this
 paper will be devoted to proving these results.
In section~\ref{samplitudes} we consider the application
of our results to on-shell amplitudes of open and closed strings.

\sectiono{The open-closed string field theory}\label{backgroundinfo}

In this section we shall  review some basic facts about open-closed string field theory. 
For more details on open-closed theory and its superstring version the 
reader may consult our recent review~\cite{Sen_Zwiebach_review} or the original
references~\cite{Zwiebach:1990qj,Zwiebach:1992bw,Zwiebach:1997fe,FarooghMoosavian:2019yke}.

\subsection{Kinetic term and propagators} \label{skinetic}

For closed strings we work in 
the Siegel gauge, thus imposing the conditions 
\be
b_0^-|\Psi_c\rangle=0, \qquad L_0^-|\Psi_c\rangle=0, \qquad b_0^+|\Psi_c\rangle=0\, .
\ee
The first two conditions are satisfied by the string field even before gauge fixing
and the last condition is the Siegel gauge condition.
For open strings the Siegel gauge takes a  
similar form
\be
b_0 \ket{\Psi_o} = 0 \, . 
\ee
 In this gauge, the quadratic part $S_2$ of the open-closed action takes
the form:
\be\label{esftactiongaugefixed} 
S_2=   \tfrac{1}{2}  \langle \Psi_c|c_0^- Q_c  |\Psi_c\rangle  + 
\tfrac{1}{2}  \langle \Psi_o|Q_o |\Psi_o\rangle =
\tfrac{1}{2}  \langle \Psi_c|c_0^- c_0^+ L_0^+ |\Psi_c\rangle  + 
\tfrac{1}{2}  \langle \Psi_o|c_0 L_0 |\Psi_o\rangle
\, ,
\ee
where $Q_c\equiv Q_B+\bar Q_B$ 
and $Q_o\equiv Q_B$ are the BRST charges
acting on the closed and open string states respectively, 
$Q_B$ and $\bar Q_B$ being the BRST charges in the holomorphic and the
anti-holomorphic sectors. 
Using the normalization conditions \refb{sl2c_overlap}, 
\refb{sl2r_overlap} and the fact that acting on closed and
open string states of  
masses $m_c$ and $m_o$ 
and momentum $k$, $L_0^+$ and $L_0$ have eigenvalues
$(k^2+m_c^2)/2$ and $(k^2+m_o^2)$ 
respectively, we see that the canonically normalized fields $\phi_c$ and $\phi_o$
will appear in the combination 
\be
\int{d^Dk\over (2\pi)^D}\phi_c(k)\, c\bar c V^c(k) \,, \qquad \hbox{and} \qquad 
\int {d^{p+1}k\over (2\pi)^{p+1}}
\phi_o(k)\, cV^o(k)\,, 
\ee 
respectively, with $V^c(z,k)$ and
$V^o(x,k)$ matter primary operators carrying momentum $k$, 
normalized as,\footnote{The
off-shell dilaton vertex operator $c\bar c V_{\rm D}(k)$ is an exception since
$V_{\rm D}(k) ={4\over \sqrt D} \eta_{\mu\nu} \p X^\mu \bar \p X^\nu\, e^{ik.X}$ is not a primary
and the operator product $V_{\rm D}(z_1,-k) V_{\rm D}(z_2, k)$ contains extra terms. However
these extra terms vanish on-shell and hence \refb{evertexnorm} still holds.}
\be  
\label{enewnorm}  
V^c(z_1,-k) V^c(z_2,k) \simeq  \  {4\over |z_1-z_2|^{4+k^2+m_c^2}}, \qquad
V^o(x_1,-k) V^o(x_2, k) \simeq  {K^{-1}\over (x_1-x_2)^{2+ 2(k^2+m_o^2)}}\, .
\ee
In that case the kinetic term of $\phi_c$ and $\phi_o$ will have the 
canonical form:
\be 
S_2  =  \int  {d^Dk\over (2\pi)^D} \bigl[ 
- \tfrac{1}{2} \phi_c(-k) (k^2 + m_c^2) \phi_c(k) \bigr] 
+ \int {d^{p+1} k \over (2\pi)^{p+1} }  \bigl[
 \, - \,  \phi_o(-k) \tfrac{1}{2} (k^2+m_o^2) \phi_o (k) \bigr]+\cdots\, ,
\ee
where $\cdots$ denote contribution from other fields.
For on-shell vertex operators, \refb{enewnorm} 
reduces to~\refb{evertexnorm}.

Let us now consider the propagators. 
The closed string kinetic operator\footnote{We work
 in the Euclidean theory in order to avoid
factors of $i$.  Given the normalization \refb{sl2c_overlap}, \refb{sl2r_overlap} 
we can see that 
the Euclidean action $S_E$ will have the correct sign if we take
$S_E = -S$,
with $S$ evaluated using the Euclidean metric.  
 Therefore the path-integral
weight factor is $e^{-S_E}=e^S$. 
The Feynman rules discussed below use this convention. 
In particular, the kinetic operator $K$ is the {\em negative}
 of the operator that appears in the quadratic term in the action.  For a scalar
 field, for example, we have $S = \int ( -\tfrac{1}{2} \phi K \phi )$ with
 $K$ the kinetic operator. With this sign convention, the interaction terms of the action $S$
 will appear in the Feynman diagram without any extra sign.}  
 here is $K_c = -c_0^- c_0^+ L_0^+$.
The closed string propagator ${\cal P}_c$ is
obtained by inverting the kinetic operator,
${\cal P}_c K_c = {\bf 1}$ on the space of states annihilated by $b_0^\pm$
and $L_0^-$. 
This gives 
\be\label{eprop}
{\cal P}_c \ = \ - b_0^+ b_0^- (L_0+\bar L_0)^{-1} \delta_{L_0,\bar L_0}\, .
\ee
The operator ${\cal P}_c$ is to be regarded as acting on the full CFT state 
space;  
this is why
we include the factor $\delta_{L_0,\bar L_0}$ to explicitly impose the 
level matching constraint for the string field. This can be written as 
\be\label{eproprep}
{\cal P}_c = \  -{1\over 2\pi} \,  b_0^+ \, b_0^- \, 
\int_0^{2\pi} \hskip-4pt d\theta\,\int_0^\infty 
\hskip-3pt ds \ 
e^{-s(L_0+\bar L_0)} e^{i\theta(L_0-\bar L_0)}\\[0.6ex]
=  \ {1\over \pi} \,  b_0 \, \bar b_0 \, \int_{|q|\le 1} \, {d^2 q\over |q|^2} q^{L_0}
\bar q^{\bar L_0}\, ,   
\ee
where we have defined 
$q\equiv e^{-s+i\theta}$ and 
$d^2q = d\theta \, ds\,  |q|^2$. We can express $d^2q$ in the form notation:
$d^2 q
 \ \equiv  \ \tfrac{i}{2} \, dq \wedge d\bar q$.
The two-form $d^2q$ is the `area' form giving a positive result acting on
a basis with the standard orientation.  It is the exact
analog of $d^2z \equiv dx \wedge dy$ for the real plane, with $z = x + iy$. 
Furthermore, using the fact that $dq\wedge d\bar q = -2i d^2 q$, the propagator
can be written as  
\be\label{epropreX}
{\cal P}_c =  \ \int_{|q|\le 1} \Bigl[  \Bigl( -{1\over 2\pi i} \Bigr) \,  b_0 \, \bar b_0 \,  \,  {dq\wedge d\bar q\over |q|^2}  \ \Bigr] \  \ q^{L_0}
\bar q^{\bar L_0}\, .   
\ee
The integration measure and ghost insertion in the propagator define the two-form 
$\Omega_{{\cal P}_c}$ shown in between brackets above 
 \be \label{eclosedprop}  
\Omega_{{\cal P}_c} \  \equiv \ - {1\over 2\pi i} \,  b_0 \,  
\bar b_0 \, {dq\wedge d\bar q\over |q|^2} 
 \, .
\ee

For the open string quadratic term in the action~\refb{esftactiongaugefixed}, 
the kinetic operator is $K_o= - c_0 L_0$. 
Therefore the open string 
propagator ${\cal P}_o$, satisfying ${\cal P}_o  K_o= {\bf 1}$ on the
space of states annihilated by $b_0$, 
 is given by 
\be \label{eopenprop}
{\cal P}_o = -b_0\, L_0^{-1} = - b_0 \int_0^1 {dq_o \over q_o} \, q_o^{L_0}\, 
=  \int_0^1 \Bigl[ (-b_0)  {dq_o \over q_o} \Bigr] \, q_o^{L_0}\, ,
\ee 
where we used a real $q_o$ as integration variable. 
The integration measure and ghost insertion in the propagator define the one-form 
$\Omega_{{\cal P}_o}$ shown in between brackets above 
 \be \label{eopenpropX}  
\Omega_{{\cal P}_o} \ = \ - b_0 \, {dq_o\over q_o} 
 \, .
\ee

\subsection{Interaction vertices} 

For {\em off-shell} 
 closed string states $A_1^c,\cdots, A_{n_c}^c$ and open string states
$A_1^o,\cdots, A_{n_o}^o$, we  define the interaction vertices 
\ben 
\label{e549newsft}
&&\hskip-30pt  \{A^c_1,\cdots, A^c_{n_c};A^o_1,\cdots, A^o_{n_o}\}_{g,b}\nonumber \\
&=& \  
 g_s^{-\chi_{g,b,n_c,n_o}} \, N_{g,b,n_c,n_o}\, 
  \int_{\VV_{g,b,n_c,n_o}}    
\hskip-20pt \Hom^{(g,b,n_c,n_o)}
(A^c_1,\cdots, A^c_{n_c}; A^o_1,\cdots, A^o_{n_o})  \, .
\een
This is similar to \refb{e549new}, but there are two differences. First, since the vertex operators
are off-shell, the result depends on the choice of local coordinates at the punctures.
Second, instead of integrating over the full moduli space $\MM_{g,b,n_c,n_o}$,
we integrate over a subspace $\VV_{g,b,n_c,n_o}$ of the moduli space. The choice
of local coordinates at the punctures and the subspaces $\VV_{g,b,n_c,n_o}$ are correlated
in a way that will
be explained below. 

For the one-point function of off-shell closed strings, the 
generalization of the result given 
in~\refb{eckv} and~\refb{e14} 
must specify the local coordinate $w$ in the disk, with the state inserted at $w=0$.  
We do this by demanding that the disk boundary is at $|w|=e^\Lambda$,
with
$\Lambda>0$ the same constant that appeared in \refb{ebinsertX}. 
Then we  have,
\be  
\label{eb14X}
\{A^c\}_{0,1} = N_{0,1,1,0}\, \Hom^{(0,1,1,0)}(A^c) = -{1\over 2\pi i}\, 
N_{0,1,1,0}\, \langle B| e^{-\Lambda L_0^+} c_0^- |A^c\rangle
= -{1\over 2\pi i}\, 
N_{0,1,1,0}\, \langle A^c| e^{-\Lambda L_0^+} c_0^- |B\rangle\,  . 
\ee
In arriving at the last expression, two minus signs have cancelled. 
One of them comes from
having to move $c_0^-$ through the Grassmann odd vertex operator $B$ and the other
comes from the BPZ conjugation of $c_0^-$.

In terms of the quantities defined above, 
the interaction terms in the string field theory action can be written as,
\be
\label{inter-terms-ocsft} 
S_{\rm int} =  \sum_{g,b,n_c,n_o} {1\over n_c!\, n_o!} \, 
\{ \Psi_c^{n_c}; \Psi_o^{n_o}\}_{g,b}\, . 
\ee
The open-closed string amplitudes to a given order in $g_s$
are now constructed in the usual manner by summing over all Feynman diagrams that
contribute to the amplitude to that order.

\subsection{Surface interpretation of Feynman diagrams} \label{s3.3}

For given $g,b$, the simplest Feynman diagram that contributes to an 
amplitude with $n_c$ closed strings and $n_o$ open strings is the one
that uses the interaction vertex proportional to $\{ \Psi_c^{n_c}; \Psi_o^{n_o}\}_{g,b}$
without any propagator. This has the same form as the amplitude given in 
\refb{e549new}, except that the integration over the moduli runs over the subspace
$\VV_{g,b,n_c,n_o}$ instead of the full moduli space $\MM_{g,b,n_c,n_o}$.

Next consider 
a Feynman diagram where a pair of interaction vertices are connected by a 
closed string 
propagator. For each of the interaction vertices we have one puncture that connects to
the internal propagator. Let $w_1$ and $w_2$ denote the local coordinates at these punctures.
One can argue that connecting two interaction vertices
by the $q^{L_0}
\bar q^{\bar L_0}$ term in the
propagator \refb{epropreX} is equivalent to  gluing together the
two closed string punctures via the identification
\be
w_1 w_2 = q  \,, \ \ \ |q| \leq 1\, .
\ee
This joins the Riemann surfaces associated with the individual interaction vertices into
a single Riemann surface 
by gluing, for example,  
 the curves $|w_1|=|q|^{1/2}$ and $|w_2|=|q|^{1/2}$ to each other 
and gives a geometric interpretation of the Feynman diagram.
The main ingredient  in the argument involves 
inserting a complete set of states along a gluing circle and using the fact that the BPZ
conjugation takes $w$ to $1/w$. The moduli of the original Riemann surface(s),
together with $q,\bar q$, form the moduli of the Riemann surfaces obtained after
the gluing.

Let us now compare the $b_0$ insertions arising from the propagator 
\refb{epropreX} with the
$\BB[\p/\p u]$ insertions that we expect in the integrand of the amplitude
associated with the Riemann surface obtained after gluing.
For the above gluing we have
$w_1 = F (w_2, q)$ with $ F (w_2, q)=q/w_2$ 
and the derivatives
\be  
{\partial F \over \partial q} = {1\over w_2}=
{w_1\over q}  \,, \ \ {\partial \bar F \over \partial q} = 0\, ; \ \ \ \  {\partial F \over \partial \bar q} = 0 \,, \ \ {\partial \bar F \over \partial \bar q} = {\bar w_1\over \bar q} \,.  
\ee 
This means that the canonical form $\widehat\Omega_c$  
arising from the gluing operation is
\be
\label{canonicalformgluing} 
\widehat\Omega_c = \BB\Bigl[ {\partial\over \partial q} \Bigr]\,  \BB \Bigl[ {\partial\over \partial \bar q} \Bigr]\,  dq \wedge d\bar q  
={1\over q} \ointclockwise w_1 b(w_1) dw_1 \cdot  {1\over \bar q} 
\ointclockwise \bar w_1 \bar 
b(\bar w_1) d\bar w_1  \,  dq\wedge d\bar q
= {b_0 \over q} \, {\bar b_0\over \bar q}  \ dq \wedge d\bar q\,,
\ee
where the origin of the clockwise contour is the same as in \refb{e224n}.
Comparing this canonical form 
with the propagator two-form $\Omega_{{\cal P}_c}$
 in~\refb{eclosedprop} we
see that they are related as follows:
\be
\label{cs-prop_vs_can}
\Omega_{{\cal P}_c}  \ = \ \AL \, \Hom_c\,,  \ \ \hbox{with} \ \ \  \AL = - {1\over 2\pi i} \,.   
\ee

\bigskip

Similarly, the effect of connecting 
two open string punctures  by an open string propagator
amounts to  
identifying local coordinates $w_1$ and $w_2$ around the open string punctures via
\be\label{e419}
w_1 w_2=-q_o\, ,  \ \ \  q_o \in [0,1]\,,
\ee
with the extra minus sign arising due to the fact that for open strings, BPZ 
conjugation
take $w$ to $-1/w$.
Identifying $w_1$ as $\sigma_s$ and $w_2$ as $\tau_s$ 
in the language used for the insertion~\refb{e233open}, we have 
$w_1 = G ( w_2 , q_o)$,  with $G(w_2,q_o)=-q_o/w_2$,
 ${\partial G\over \partial q_o} = w_1/q_o$.  
This gives the canonical form in the  
open string case:
\be  
\Hom_o\ = \ \BB\Bigl[ {\partial \over \partial q_o}  \bigr] \, dq_o  \, ,\ee
where
\be
\BB\left[{\p\over \p q_o}\right] = \int dw_1 \, b(w_1) \,  {w_1\over q_o} 
+\int  d\bar w_1 \, \bar b(\bar w_1) \, {\bar w_1\over q_o}={1\over q_o}
\ointclockwise_C dw_1 b(w_1) w_1\, .
\ee
Identifying 
$\ointclockwise_C dw_1 b(w_1) w_1= -b_0$, 
we get, 
\be\label{eminus}  
\BB\left[{\p\over \p q_o}\right] =-{1\over q_o} \, b_0, \qquad 
\Hom_o\ = -  b_0\, {dq_o\over q_o} \, . 
\ee
Comparing this with 
the open-string propagator form~\refb{eopenpropX}
we get the relation: 
\be
\label{open-canon-vs-norm} 
\Omega_{{\cal P}_o} = \, 
\Hom_o\,. 
\ee
This shows that for open strings the propagator and canonical forms are exactly the same.

This analysis can be extended
to general Feynman diagrams with multiple propagators,
and shows that the contribution from each Feynman diagram takes the form of the
world-sheet amplitude \refb{e549new} except for two issues. First the region of integration
over the parameters of the moduli space is determined from the regions $\VV_{g,b,n_c,n_o}$
in the interaction vertices of the diagram and the ranges $|q_c|\le 1$, 
$0\le q_o\le 1$ associated  
with each propagator, and it is not {\it a priori} clear what region of the moduli space they cover. 
One can show that in a gauge invariant string field theory the regions $\VV_{g,b,n_c,n_o}$ are
such that the union of the regions of the moduli space associated with different  Feynman diagrams give the full moduli space.  Examples of 
the regions $\VV_{g,b,n_c,n_o}$ and the choice of local coordinates
at the punctures can be provided
using minimal area metrics~\cite{Zwiebach:1997fe} or hyperbolic geometry~\cite{Costello:2019fuh,Cho:2019anu}. 
Second,  
the normalization
constants associated with different  Feynman diagrams contributing to an amplitude must agree
with each other and with the expected normalization given in \refb{e549new}. 
In the next two sections we shall use these conditions to determine the normalization constants
$N_{g,b,n_c,n_o}$.
To the best of our knowledge, this has not been done before.

 \sectiono{Recursion relations for the normalization constants}
   
\label{NormForm}

In this section we will use the requirement that gluing string field theory interaction
vertices by propagators generates  
the amplitude with correct normalization
to derive a set of recursion relations   
for
the constants $N_{g,b,n_c,n_o}$. 
In subsection \ref{s4.1} we describe the general procedure for deriving the constraints
up to sign factors. In subsections \ref{s4.2} and \ref{s4.3} we fix the signs in 
special cases. 
In section~\ref{sanalysis} we shall
use these relations to determine $N_{g,b,n_c,n_o}$.

\subsection{Recursion relations for normalization coefficients  up to signs} \label{s4.1}

The analysis of section \ref{s3.3} shows that if the 
antighost insertions in the propagators had
used the canonical forms $\Hom_c$ and $\Hom_o$, then the 
correlation function obtained after gluing a pair of $\Hom^{(g_i,b_i,n_{ci}, n_{oi})}$
via a propagator would have generated the form $\Hom^{(g,b,n_c,n_o)}$ on the moduli
space of the
Riemann surfaces obtained after gluing, up to signs that may arise from the rearrangement
of the $\BB$'s and the unintegrated 
open string vertex operators. Then $N_{g,b,n_c,n_o}$ would have
been equal to the product of $N_{g_i,b_i,n_{ci}, n_{oi}}$'s up to signs. The fact that the
antighost insertions in the closed string propagator
give $\AL\Hom_c$ leads to an extra factor of $\AL$ multiplying the 
product of $N_{g_i,b_i,n_{ci}, n_{oi}}$'s
whenever we glue two punctures using a closed string propagator. 
No such extra factor appears when we connect
a pair of interaction vertices by an open string propagator.
The same 
consideration applies to the case when we glue two punctures on the same Riemann surface
via a propagator.

\begin{figure}[h]
	\centering
\epsfysize=8.0cm
\epsfbox{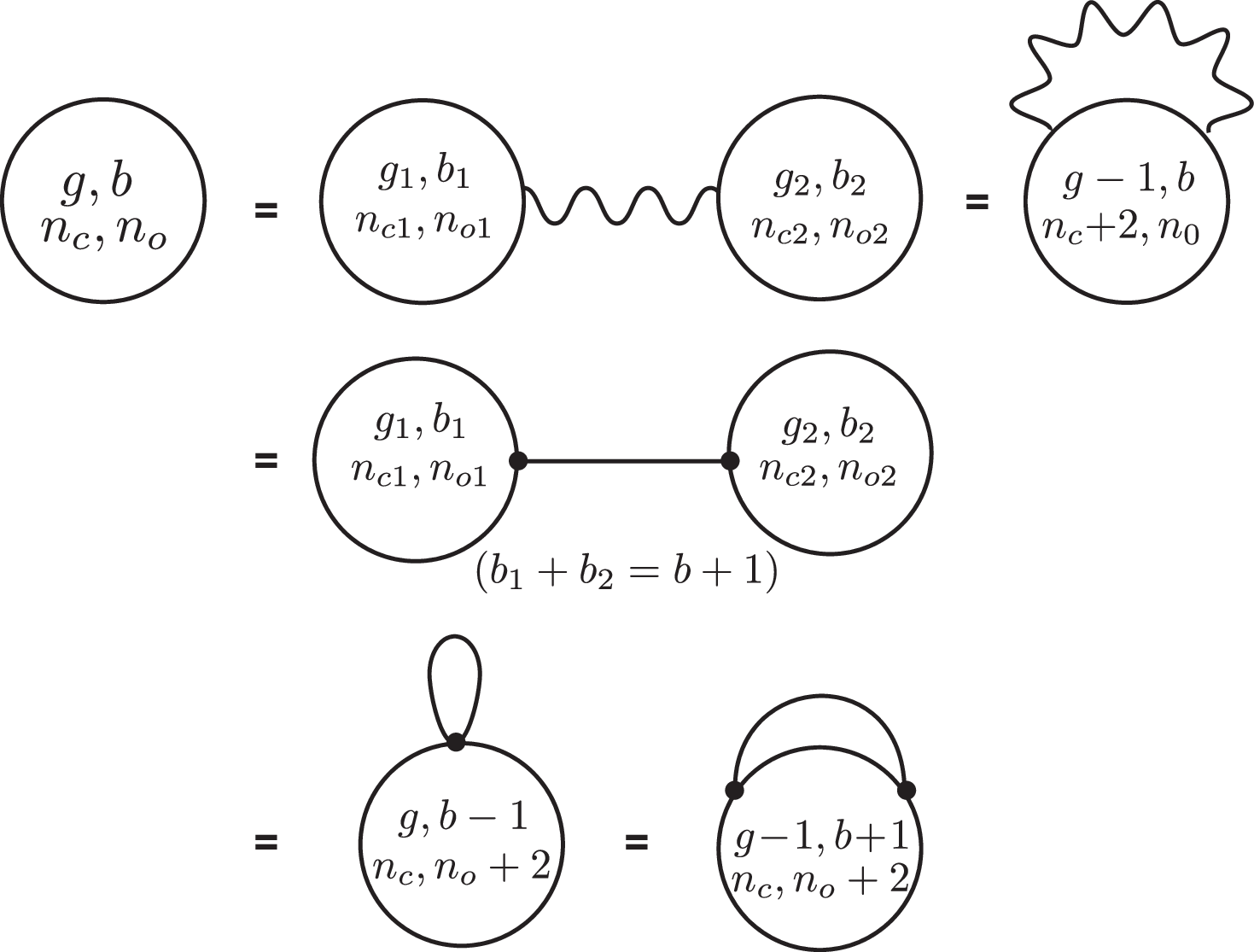}
	\caption{\small 
	The surfaces in the open-closed moduli spaces can be built 
	 in five different ways, all leading to amplitudes that must have identical normalizations.   The first two use closed string 
	 propagators denoted by wavy lines, 
	 the last three use open string
	 propagators denoted by continuous lines. 
	 Heavy dots denote boundary components, and
	 external open and closed string states are not shown. 
	In the  
	diagrams we have $g=g_1+g_2$ and $b=b_1+b_2$ unless  stated
	otherwise.} 
	\label{xhy}
\end{figure}

We now note that a set of surfaces in
$\MM_{g,b,n_c,n_o}$ of the top dimensionality 
can, in general,
 be built in five
different ways using sets in lower-dimension moduli spaces and single propagator.
This was noted in writing the geometric version of the Batalin Vilkovisky master equation in 
open-closed string field theory~\cite{Zwiebach:1997fe}. 
As shown in Figure~\ref{xhy} these five ways are
\begin{enumerate} 
\item  Joining  sets of type $(g_1, b_1, n_{c1}, n_{o1})$ and $(g_2, b_2, n_{c2}, n_{o2})$
via a closed string propagator, with $g= g_1 + g_2$,  $b= b_1 + b_2$, $ n_c = n_{c1}+n_{c2} - 2$, and $n_o = n_{o1} + n_{o2}$. 

\item  Joining via a closed string propagator
 two punctures in a set of type 
$(g-1, b, n_{c} + 2 , n_{o})$.  

\item  Joining  via an open string propagator 
sets of type $(g_1, b_1, n_{c1}, n_{o1})$ and $(g_2, b_2, n_{c2}, n_{o2})$, with $g= g_1 + g_2$,  $b= b_1 + b_2-1$, $ n_c = n_{c1}+n_{c2} $, and $n_o = n_{o1} + n_{o2}-2$. 

\item  Joining via an open string propagator two open string punctures lying on the same boundary component
 in a set of type 
$(g, b-1, n_{c} , n_{o}+2 )$. 

\item  Joining via an open string propagator two open string punctures lying on 
different boundary components
 in a set of type 
$(g-1, b+1, n_{c} , n_{o}+2 )$.

\end{enumerate} 

\medskip 
\noindent

The argument given at the beginning of this subsection now establishes the  following
five constraints on the normalization constants $N_{g,b,n_c,n_o}$:
\be
 \label{ecombinedX}
\begin{split} 
& \  N_{g_1+g_2,\, b_1+b_2,\,  n_{c1}+n_{c2}-2,\,  n_{o1}+n_{o2}}\  \Hom^{(g_1+g_2,\, b_1+b_2,\,  n_{c1}+n_{c2}-2,\,  n_{o1}+n_{o2})}  \\
& \  \ \ \  =   \, \AL \, N_{g_1,b_1,n_{c1},n_{o1}} N_{g_2,b_2,n_{c2},n_{o2}}\ \Hom_c \cup\Hom^{(g_1,b_1,n_{c1},n_{o1})}  \cup \Hom^{(g_2,b_2,n_{c2},n_{o2})} \,,  \\[2.0ex]
& N_{g,\, b,\, n_c,\, n_o} \ \Hom^{(g,\, b,\, n_c,\, n_o)} =   \ \, \AL\, N_{g-1,b,n_c + 2,n_o} \
\Hom_c \cup \Hom^{(g-1,b,n_c + 2,n_o )} \,, 
\\[2.0ex]
& N_{g_1+g_2,\, b_1+b_2-1,\, n_{c1}+n_{c2},\, n_{o1}+n_{o2}-2}\ 
\Hom^{(g_1+g_2,\, b_1+b_2-1,\, n_{c1}+n_{c2},\, n_{o1}+n_{o2}-2 )}
 \\
& \ \ \  =   \ \, 
N_{g_1,b_1,n_{c1},n_{o1}} N_{g_2,b_2,n_{c2},n_{o2}} \   \Hom_o \cup 
\Hom^{(g_1,b_1,n_{c1},n_{o1})} \cup  \Hom^{( g_2,b_2,n_{c2},n_{o2}) } \,,   \\[2.0ex]
&  N_{g,b,n_c,n_o}\ \Hom^{(g,b,n_c,n_o)}  =  \ \,  N_{g,b-1,n_c,n_o+2}\ 
\Hom_o \cup \Hom^{(g,b-1,n_c,n_o+2)} \,, \\[2.0ex]
&  N_{g,b,n_c,n_o} \ \Hom^{(g,b,n_c,n_o)}   
=   \ \,  N_{g-1,b+1,\, n_c,\, n_o+2 } \ \Hom_o \cup \Hom^{(g-1,b+1,\, n_c,\, n_o+2)} \, ,
\end{split}
\ee
where we use the 
 symbol $\cup$ to denote the joining of canonical forms to construct a 
 canonical form of larger degree. 
The first two relations arise from the closed string propagator
 joining punctures on two surfaces or on a single surface, respectively.   
 We include the form $\AL\,\Hom_c$
 for the closed string propagator. 
 The third relation is from the open
string propagator joining punctures on two different sets of surfaces.  
The fourth relation is from the open string propagator joining two punctures
on the same boundary component,  and
the final relation is from the open string propagator joining two punctures 
on different boundary components of the same surface.   

In each equation above, the canonical form on the left-hand side coincides, up to a sign, with
the canonical form on the right-hand side. The signs arise from the ambiguity in the 
ordering of the various insertions in the correlators.   Thus, in principle we 
have the relations 
\be  
 \label{ecombinedXX}
\begin{split} 
 \   \Hom^{(g_1+g_2,\, b_1+b_2,\,  n_{c1}+n_{c2}-2,\,  n_{o1}+n_{o2})}  
 \    &\sim    \ \Hom_c \cup\Hom^{(g_1,b_1,n_{c1},n_{o1})}  \cup \Hom^{(g_2,b_2,n_{c2},n_{o2})}\, ,  \\[2.0ex]
  \Hom^{(g,\, b,\, n_c,\, n_o)} \  & \sim   \
\Hom_c \cup \Hom^{(g-1,b,n_c + 2,n_o )} \, ,
\\[2.0ex]
\Hom^{(g_1+g_2,\, b_1+b_2-1,\, n_{c1}+n_{c2},\, n_{o1}+n_{o2}-2 )} 
\   & \sim    \   \Hom_o \cup 
\Hom^{(g_1,b_1,n_{c1},n_{o1})} \cup  \Hom^{( g_2,b_2,n_{c2},n_{o2}) }\, ,   \\[2.0ex]
  \Hom^{(g,b,n_c,n_o)} \ &\sim   \ 
\Hom_o \cup \Hom^{(g,b-1,n_c,n_o+2)}\, ,  \\[2.0ex]
  \Hom^{(g,b,n_c,n_o)}   
\ & \sim   \    \Hom_o \cup \Hom^{(g-1,b+1,\, n_c,\, n_o+2)} \, ,
\end{split}
\ee
where $\sim$ denotes equality up to sign.
If the forms only glue properly up to signs, those signs must be corrected
by the $N$ coefficients in order to give the equalities in~\refb{ecombinedX}:
\be
 \label{ecombined}
\begin{split} 
 N_{g_1+g_2,\, b_1+b_2,\,  n_{c1}+n_{c2}-2,\,  n_{o1}+n_{o2}}
\sim & \  \, \AL \, N_{g_1,b_1,n_{c1},n_{o1}} N_{g_2,b_2,n_{c2},n_{o2}},  \\[0.5ex]
 N_{g,\, b,\, n_c,\, n_o} \sim & \ \, \AL\, N_{g-1,b,n_c + 2,n_o}, \\[1.0ex]
 N_{g_1+g_2,\, b_1+b_2-1,\, n_{c1}+n_{c2},\, n_{o1}+n_{o2}-2}
\sim & \ \, 
N_{g_1,b_1,n_{c1},n_{o1}} N_{g_2,b_2,n_{c2},n_{o2}},  \\[0.5ex]
 N_{g,b,n_c,n_o} \sim & \ \,  N_{g,b-1,n_c,n_o+2}, \\[0.5ex]
 N_{g,b,n_c,n_o}   
\sim  & \ \,  N_{g-1,b+1,\, n_c,\, n_o+2 } \, . 
\end{split}
\ee

\medskip
To proceed further with the analysis,
let us now define primed canonical forms.  For this we first define 
primed correlation functions as
\be
\label{Kcorrprimed} 
\langle \cdots \,  \rangle \equiv  K^{b}  
\langle \cdots\,  \rangle' \, .
\ee
This is 
more than a definition.  In writing this expression, we claim that
the primed correlator has no dependence on $K$ 
{\em when}
computed in the 
closed string channel by adding boundaries to a purely closed string 
amplitude;  
all the dependence is in
the factor $K^b$  that was brought into the open.  This follows because each time we
add a boundary it can be obtained by gluing into a surface, via a closed string
propagator,  a disk with a single closed string puncture and a number of open string punctures.
Such a disk amplitude is 
proportional to $K$ due to \refb{eopencorr}, 
and thus the inclusion of $b$ 
boundaries will give the factor $K^b$ shown above. 
If we now define $\Hom^{\prime(g,b,n_c,n_o)}$ by replacing the usual correlation
function by the primed correlation function on the right hand side of \refb{edefOmegaOpen},
we recover the definition given in \refb{edefHomp}.
Additionally, we will rescale the open string field $\Psi_o$ 
 by writing
 \be
\Psi_o = {1\over \sqrt{K}} \psi_o \,,
\ee
so that the kinetic term for the open string field takes the form:
\be
S_{\rm kin}^o =
 \tfrac{1}{2}  \langle \Psi_o|Q_o|\Psi_o\rangle = \,  
\tfrac{1}{2} \langle \psi_o|Q_o|\psi_o\rangle'\, .
\ee
Then we have the relation
\be
\label{roeivdj}
\Hom^{(g,b,n_c,n_o)} (\Psi_c^{n_c};\Psi_o^{n_o}) = K^{b - {1\over 2} n_o}\ 
\Hom^{\prime (g,b,n_c,n_o)}
 (\Psi_c^{n_c};\psi_o^{n_o})\, .
\ee
In these variables the interaction terms~\refb{inter-terms-ocsft} 
of the open closed string field theory may be written as
\ben \label{enewint}
S_{\rm int} &=& \hskip-8pt \sum_{g,b,n_c,n_o} {1\over n_c!\, n_o!} \, 
\{ \Psi_c^{n_c}; \Psi_o^{n_o}\}_{g,b}\nonumber \\
&=& \hskip-8pt \sum_{g,b,n_c,n_o} {1\over n_c!\, n_o!} \, g_s^{-\chi_{g,b,n_c,n_o}} 
N_{g,b,n_c,n_o} 
 K^{b - {1\over 2} n_o} 
\int_{\VV_{g,b,n_c,n_o} } \hskip-8pt \Hom^{\prime (g,b,n_c,n_o)}
 (\Psi_c^{n_c};\psi_o^{n_o})\, , 
\een
where we used the interaction vertices in~\refb{e549newsft} and the primed forms
in~\refb{roeivdj}. 
It will be understood in the rest of this subsection 
that the arguments of $\Hom$ will be $\Psi_c$ and $\Psi_o$, while the
arguments of $\Hom'$ will be $\Psi_c$ and $\psi_o$.

Note that in the new variables $\Psi_c,\psi_o$, the 
propagator 
forms remain unchanged
since  the kinetic term of $\psi_o$ 
in terms of the primed correlation function
has the same form as the kinetic term of $\Psi_o$ in terms of the
unprimed correlation function.  The main difference is the $K^{b - {1\over 2}n_o}$ factor
in the interaction term \refb{enewint}. As a result \refb{ecombinedXX} is modified to:
\be 
 \label{primed-ecombinedXX}
\begin{split} 
 \   \Hom'^{(g_1+g_2,\, b_1+b_2,\,  n_{c1}+n_{c2}-2,\,  n_{o1}+n_{o2})}  
 \    &\sim    \ \Hom_c \cup\Hom'^{(g_1,b_1,n_{c1},n_{o1})}  \cup \Hom'^{(g_2,b_2,n_{c2},n_{o2})}\, ,  \\[2.0ex]
  \Hom'^{(g,\, b,\, n_c,\, n_o)} \  & \sim   \
\Hom_c \cup \Hom'^{(g-1,b,n_c + 2,n_o )} \, ,
\\[2.0ex]
\Hom'^{(g_1+g_2,\, b_1+b_2-1,\, n_{c1}+n_{c2},\, n_{o1}+n_{o2}-2 )} 
\   & \sim    \   \Hom_o \cup 
\Hom'^{(g_1,b_1,n_{c1},n_{o1})} \cup  \Hom'^{( g_2,b_2,n_{c2},n_{o2}) }\, ,   \\[2.0ex]
  \Hom'^{(g,b,n_c,n_o)} \ &\sim   \  K^{-2} \  
\Hom_o \cup \Hom'^{(g,b-1,n_c,n_o+2)}\, , \\[2.0ex]
  \Hom'^{(g,b,n_c,n_o)}   
\ & \sim   \    \Hom_o \cup \Hom'^{(g-1,b+1,\, n_c,\, n_o+2)} \, .
\end{split}
\ee
In all the relations, except the fourth, the factors of $K$ cancel out.
This has some implications.  Since $K$ is unknown and to be determined,
only the fourth condition, that of joining of open strings within a boundary,
has the ability to determine $K$.  Moreover, we can exploit this to our
convenience.  Since $K$ is undetermined, 
we can actually demand that $K$
is such that this fourth equation holds {\em strictly}
\be
\label{4thstrictly}
 \Hom'^{(g,b,n_c,n_o)} \ =    \  K^{-2} \  
\Hom_o \cup \Hom'^{(g,b-1,n_c,n_o+2)} \, . 
\ee
This fixes the value of $K$, as we will discuss later.  But now, passing back to
canonical forms, we will also have their equality,\footnote{One 
might wonder whether
by using a single constant $K$ one can satisfy \refb{4thstrictly} for all $g,b,n_c,n_o$.
We shall see below \refb{esolrecnewNN} 
that once we satisfy \refb{4thstrictly} for $(g,b,n_c,n_o)= (0,2,0,1)$, 
it holds for all other $g,b,n_c,n_o$.}
\be
\label{4thstrictlyX}
 \Hom^{(g,b,n_c,n_o)} \ =    \ \  
\Hom_o \cup \Hom^{(g,b-1,n_c,n_o+2)} \, .
\ee
As a final consequence, the associated equation for the $N$ coefficients
will also hold strictly:
\be
\label{XYZ}
 N_{g,b,n_c,n_o} \  =  \   N_{g,b-1,n_c,n_o+2} \, . 
 \ee

In the next two subsections we shall show that for the prescription
discussed for $\Hom^{(g,b,n_c,n_o)}$ in section \ref{sconv}, 
 a few of the other equations in \refb{ecombined} also work  
 without extra signs,
and those suffice to determine all the $N_{g,b,n_c,n_o}$'s.

\subsection{Gluing with closed string propagators} \label{s4.2}

In this subsection we shall resolve the sign ambiguities that arise when we join
two interaction vertices by a closed string propagator. Since  the
insertion of $\BB[\p/\p q]\BB[\p/\p{\bar q}]$ associated with the propagator
is a Grassmann even operator, 
there is no sign ambiguity involving their position inside the correlator. The
integration measure $dq\wedge d\bar q$ is also unambiguous since we can take
$q=q_x+iq_y$, write $dq\wedge d\bar q=-2\, i\, dq_x\wedge dq_y$ and take
$\int dq_x\wedge dq_y$ to be positive. Also in our prescription for
$\Hom^{(g,b,n_c,n_o)}$
the insertions associated with the 
moduli of Riemann surfaces without boundaries or open
strings always come in Grassmann even combinations 
$\BB_u \BB_{\bar u}$
and insertion of additional boundaries
and open strings following prescriptions P1 and P2 
also come in Grassmann even combinations.
Therefore as long as the Riemann surfaces associated with the interaction vertices 
can be built using prescriptions P1 and/or P2, 
there is no sign ambiguity, and we have 
\be\label{eclosedseparating2}
N_{g_1+g_2,b_1+b_2,n_{c1}+n_{c2}-2,n_{o1}+n_{o2}}
= \, \AL \, N_{g_1,b_1,n_{c1},n_{o1}} N_{g_2,b_2,n_{c2},n_{o2}}\, ,
\ee
\be \label{eproptwo}
N_{g+1,b,n_c-2,n_o} =\AL\, N_{g,b,n_c,n_o} \, . 
\ee
 
However, sign subtleties
in \refb{eclosedseparating2} and \refb{eproptwo}
 can arise when
one (or both) of
the participating surfaces  contain an unintegrated
open string vertex operator or a fixed size boundary. 
There are two basic 
surfaces of this type to be considered: 
a disk with one closed
string and a disk 
with one closed and one open string. 
The addition of more open string vertex operators can be 
carried out using Grassmann even integrated vertex operators following 
P2 and does not alter the analysis.
Note that these surfaces requiring special attention 
are not relevant to the gluing 
constraint~\refb{eproptwo}, where the number of closed string
punctures on the surface must be at least two.  So this 
constraint has no sign ambiguities.  

We shall now discuss the two cases relevant to~\refb{eclosedseparating2}. 

\begin{enumerate}

\item[1.] 

First consider the case
when we use a closed string propagator to connect 
a particular  
puncture in a Riemann
surface of type $(g,b,n_c+1,n_o)$ to a Riemann surface of type $(0,1,1,0)$,
a disk with one closed string puncture.  
The result is part of 
the amplitude 
${\cal A}_{g,b+1, n_c, n_o}$ 
for a surface with one
more boundary and one less closed string puncture. 
As in \refb{e428a}, 
let us denote by $\Sigma$ the state
such that $\langle \Sigma| \psi_c\rangle$ gives 
the expression for $\Hom^{(g,b,n_c+1,n_o)}$
with the ghost number two state $\psi_c$   
inserted
into the $(n_c+1)$-th puncture. 
Since the
boundary state $|B\rangle$ 
automatically satisfies the $L_0=\bar L_0$
condition, the projection $\delta_{L_0,\bar L_0}$ in the closed string propagator 
\refb{eprop} acts trivially and, using \refb{eb14X}, 
the resulting sewn 
amplitude takes the form:
\be\label{e434a} 
{\cal A}_{g,b+1, n_c, n_o} ^\inc = 
g_s^{-\chi_{g,b+1, n_c, n_o}}\  (\AL 
\, N_{0,1,1,0})\, N_{g,b,n_c+1,n_o}\, \int
\langle\Sigma|(-b_0^+b_0^- ) \int_0^1 {d\Qr\over \Qr} 
\Qr^{L_0^+}   
e^{-\Lambda L_0^+ }  
c_0^- | B\rangle \, ,
\ee
where the first integral runs over  
$\MM_{g,b,n_c+1,n_o}$ and we used the fact that the Euler numbers of the surfaces
add under
gluing.   The superscript $\inc$ on $\AAA$ 
denotes that the right hand side of 
\refb{e434a} gives
part of the contribution to ${\cal A}_{g,b+1, n_c, n_o}$.
Since $b_0^-$ kills the boundary state, 
we are left with
\ben\label{e425}  
{\cal A}_{g,b+1, n_c, n_o}^\inc &=&  - g_s^{-\chi_{g,b+1, n_c, n_o}}\  \AL
\, N_{0,1,1,0}\, N_{g,b,n_c+1,n_o}\, \int \langle\Sigma| \int_0^1 
{d\Qr\over \Qr} (e^{-\Lambda} \Qr)^{L_0^+} 
b_0^+   
|B\rangle \nonumber \\
&=& 
g_s^{-\chi_{g,b+1, n_c, n_o}}\  (\AL 
\, N_{0,1,1,0})\, N_{g,b,n_c+1,n_o}\, \int \Hom^{(g,b+1,n_c,n_o)}
\, ,
\een
where in the last step we used \refb{esewtoA}. 
On the other hand, this  amplitude
must be given by the integral of 
$g_s^{-\chi_{g,b+1, n_c, n_o}} N_{g, b+1, n_c, n_o}\, \Hom^{(g,b+1,n_c,n_o)}$.
This gives,
\be \label{e433}
N_{g,b+1,n_c,n_o} 
=\, \AL\, N_{g,b,n_c + 1,n_o} \, N_{0,1,1,0}\, . 
\ee
This shows that \refb{eclosedseparating2} holds 
also for the gluing of the 
Riemann surfaces of
type $(0,1,1,0)$.  
It also justifies our choice of the 
Grassmann even
one-form state 
$\ket{\cc}$ in~\refb{ebinsertX} to be inserted in the prescription
to create a boundary.  

\item[2.]
Second we consider the case where we use a closed string propagator
to connect a closed string  
puncture on a 
Riemann surface $\Sigma$ of type $(g,b, n_c, n_o)$ 
to a surface of type $(0,1,1,1)$, namely
a disk $D$ 
 with one open string and one closed string puncture.
Let $|\Sigma\rangle$ denote the state such that
if we insert the state $\psi_c$ at the special puncture of $\Sigma$
that has to be connected to the propagator, 
then we have
$\Hom^{(g,b,n_c,n_o)}(\psi_c, \cdots)   
= \langle \Sigma|\psi_c\rangle =\langle \psi_c|\Sigma\rangle$.
Using the propagator form $\Omega_{{\cal P}_c}$ in  
\refb{eclosedprop}, it follows that the  
amplitude ${\cal A}^{\inc}_{g, b+1 , n_c-1, n_o+1}$ 
obtained
by gluing can be regarded as the result of placing at the
center of the disk $D$, with its open string puncture at the boundary, 
the closed string 
state  $(g_s^{-\chi_{g,b, n_c, n_o}} 
N_{g,b,n_c,n_o}{\cal O}_\Sigma)$ with  ${\cal O}_\Sigma$ given by 
\be\label{eb.3}
|{\cal O}_\Sigma\rangle  = \   
\AL \, \int_{|q|\le 1} {dq\over q} \wedge {d\bar q\over \bar q}\, q^{L_0} \bar q^{\bar L_0}
b_0\, \bar b_0\, |\Sigma\rangle\, .
\ee
Geometrically this corresponds to identifying 
the local coordinate $z$ at the closed string puncture on the disk 
and the local coordinate $w$ at the puncture on 
$\Sigma$ via
\be
zw = q\, .
\ee
The geometry of the situation and some of the following discussion
refers to Figure~\ref{FF3FF}. 
For simplicity, let us take $z$ to be the coordinate in which the 
boundary of $D$
 is at $|z|=1$.
Writing $q=r\, e^{i\theta}$, we can express the state
$|{\cal O}_\Sigma\rangle$ in~\refb{eb.3} as
\be
|{\cal O}_\Sigma\rangle = \   
\AL \, \int_0^1 {dr\over r} (-2i)\int_0^{2\pi} d\theta\, \left(
\tfrac{1}{2}  b_0^- b_0^+\right) r^{L_0^+} e^{i\theta L_0^-}
|\Sigma\rangle\, . 
\ee
Let us suppose that in the coordinate system $z$ 
the open string vertex operator $cV^o$ is inserted at $z=1$. 
By the rule for disk amplitudes with one open and one closed
insertion in~\refb{esignchange},   
the amplitude ${\cal A}^{\inc}_{g, b+1 , n_c-1, n_o+1}$ 
is then given~by 
\ben \label{ea11a} 
&& \hskip-20pt {\cal A}_{g, b+1 , n_c-1, n_o+1}^{\inc}   \ = \  
 - g_s^{-\chi_{g, b+1 , n_c-1, n_o+1}}  N_{0,1,1,1} \, N_{g,b,n_c,n_o} \, 
\bigl\langle 
cV^o (z=1)  \ {\cal O}_\Sigma  (z=0) \bigr\rangle \nonumber \\[1.3ex]
&=&\hskip-5pt  - g_s^{-\chi_{g, b+1 , n_c-1, n_o+1}} N_{0,1,1,1} \, 
N_{g,b,n_c,n_o}   \nonumber \\ && \hskip .3in
 \  \,  
\bigl\langle B|
cV^o (z=1) \, \AL \, \int_0^1 {dr\over r} (-2i)\int_0^{2\pi} d\theta\, \left(
\tfrac{1}{2}  b_0^- b_0^+\right) r^{L_0^+} e^{i\theta L_0^-}
|\Sigma\rangle\, .  
\een
We now use the result
\be
e^{-i\theta L_0^-} cV^o(z=1) 
e^{i\theta L_0^-} 
= cV^o(z=e^{-i\theta})\, ,
\ee
and the fact that $L_0^-$ annihilates $\langle B|$, to remove the
$e^{i\theta L_0^-}$ factor in \refb{ea11a} at the cost of changing the
argument of $cV^o$ to $e^{-i\theta}$.
The
$b_0^-$ contour, surrounding $z=0$ and running anti-clockwise, 
can be deformed to the boundary, effectively 
enclosing the vertex operator 
$cV^o$  
by a clockwise semicircle. Using the doubling trick 
we can express this as 
\be\, -\,  \ointclockwise 
z b(z) dz c V^o(e^{-i\theta})= 
e^{-i\theta} V^o(e^{-i\theta}) .
\ee 
The minus sign on the left-hand side arises because 
$b_0^-$ in the
 disk correlator~\refb{ea11a} appears to the right of $cV^o$ and it must be moved
to the left of $cV^o$.  A second minus sign leading to the right-hand side answer, 
appears as the contour around
$cV^o$ is clockwise.  All in all, the $b_0^-$ operator changes $cV^o(z=e^{-i\theta})$ into
$V^o(z=e^{-i\theta}) \, e^{-i\theta} $.  The amplitude \refb{ea11a} then
becomes 
\ben \label{ea.12}
{\cal A}_{g, b+1 , n_c-1, n_o+1}^{\inc}   = g_s^{-\chi} 
  N_{0,1,1,1} \, N_{g,b,n_c,n_o} \ i \AL  
  \int_0^1 \hskip-3pt {dr\over r}\hskip-3pt \int_0^{2\pi} \hskip-6pt d\theta\,   
\bigl\langle B| 
V^o (z=e^{-i\theta}) \, e^{-i\theta}  \, b_0^+
r^{L_0^+} |\Sigma\rangle\, ,
\een
where we write $\chi = \chi_{g, b+1 , n_c-1, n_o+1}$ for brevity. 
\begin{figure}[h]
	\centering
\epsfysize=6.3cm
\epsfbox{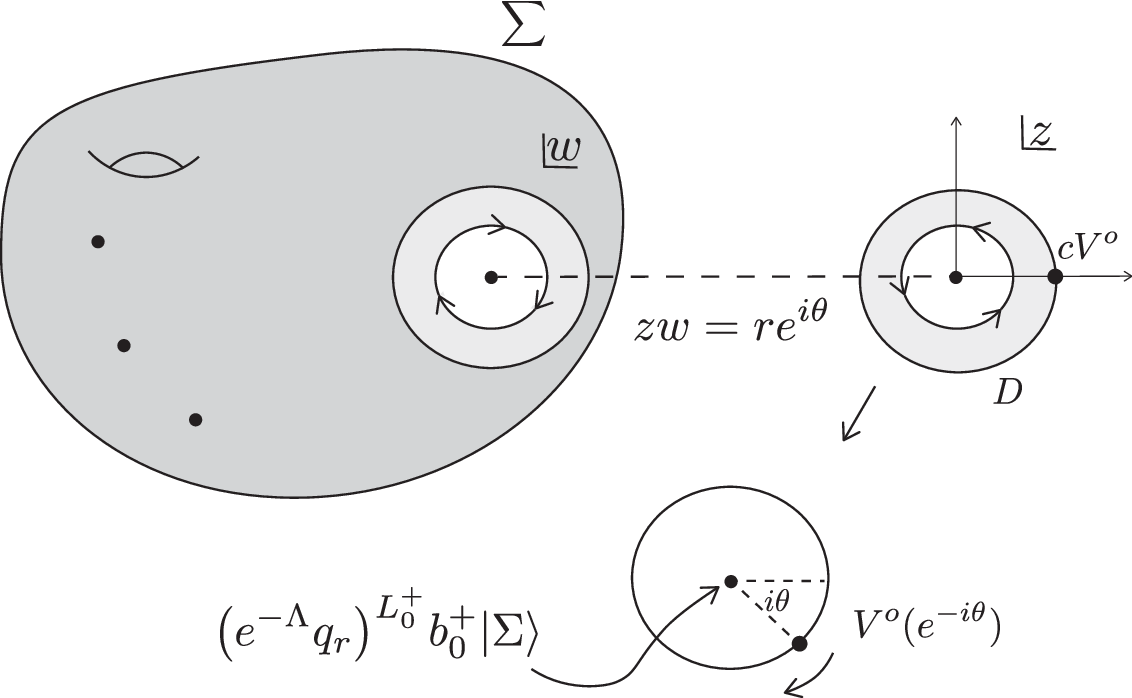}
	\caption{\small   
	The  $w=0$ puncture on $\Sigma$ is glued, via
          a closed string propagator, to a disk $|z|\leq 1$ with a 
          closed string puncture at $z=0$ and 
	an open string operator $cV^o$ at $z=1$.  The result is a surface
	$\Sigma_B$ with a boundary, 
	similar to that in figure \ref{FF2FF},  
	but including an open string insertion integrated over
	the boundary.  The gluing curves, shown in the top figure, support
a $b_0^+b_0^-$ insertion from the propagator of which the $b_0^-$ 
is made to act on $cV^o$. 
	The amplitude can also be represented (as indicated by an arrow)
	as a disk amplitude with an insertion of a closed string state at $z=0$ and a 
	moving open string insertion.} 
	\label{FF3FF}
\end{figure}
\medskip 
\noindent

 Defining $u=e^{-i\theta}$,
we can now write,
\be
\int_0^{2\pi} d\theta\,  V^o (z=e^{-i\theta}) \, e^{-i\theta} 
=i \ointclockwise du \, V^o(z=u)\, ,
\ee
where $\ointclockwise$ stands for integration along a clockwise contour since
as $\theta$ runs from $0$ to $2\pi$,
$u$ traverses the boundary in the clockwise direction. Changing the direction of 
integration produces an extra minus sign and substituting this into \refb{ea.12} we get
\be \label{eb.60pre}  
{\cal A}_{g, b+1 , n_c-1, n_o+1}^{\inc}   \ = \ 
 g_s^{-\chi}  
 N_{0,1,1,1} \, N_{g,b,n_c,n_o} \  \AL\, \bigl\langle B|  \ointop du \, V^o (u)
  \,  \int  {dr\over r} 
 \,  \, b_0^+
r^{L_0^+} |\Sigma\rangle\, ,
\ee 
where, as usual, $V^o(u)$ is the shorthand notation for $V^o(z=u)$.
Finally, we introduce a new variable $\Qr=e^{\Lambda}r$ to express \refb{eb.60pre} as 
\be\label{eb.60}   
{\cal A}_{g, b+1 , n_c-1, n_o+1}^{\inc} \,= \,   g_s^{-\chi} 
  N_{0,1,1,1} \, N_{g,b,n_c,n_o} \,   
\AL \, \left\langle B\left|\int du V^o(u) \,\int {d\Qr\over \Qr} b_0^+\,
(e^{-\Lambda}\Qr)^{L_0^+} \right|\Sigma\right\rangle\, .
\ee

Comparing \refb{eb.60} with the right hand side of \refb{esewtoA} 
we see that 
we have the correct integration measure
over $\Qr$  for the construction of $\Hom^{(g, b+1 , n_c-1, n_o)}$ 
together with an insertion of 
$\AL \int du V^o(u)$. 
Furthermore, the $u$ integral that runs counter-clockwise
along the boundary keeps the
world-sheet to the left as required in P2.
Therefore we have the expected integral that should appear
in $\int \Hom^{(g,b+1,n_c-1, n_o+1)}$ 
without any additional
sign.
Taking into account the $\AL$ factor in \refb{eb.60}, we get the relation
\be \label{eb.13a}
N_{g,b+1,n_c-1, n_o+1} = \AL \, N_{0,1,1,1} \, N_{g,b,n_c,n_o}\, .
\ee
This is again a special case of 
\refb{eclosedseparating2}.

\end{enumerate}

\subsection{Gluing with open string propagators} \label{s4.3}

By analyzing Feynman diagrams built with open string propagators, we can
get additional  constraints on the normalization
coefficients $N_{g,b,n_c,n_o}$.
As discussed earlier,  the relevant equations 
are the last three in~\refb{ecombined}.  The sign ambiguities,
encoded in the $\sim$ symbol, 
arise
because after gluing via
open string propagators surfaces whose insertions are consistent with 
the prescriptions, we have no guarantee that insertions on 
the sewn surface
are still  consistent with the prescription.
To resolve the ambiguities we need to  carefully keep track 
of the signs in the gluing
via open string propagators.
The goal in this subsection will be to resolve these ambiguities in a few cases that
will then suffice to determine the $N_{g,b,n_c,n_o}$ via a recursive procedure.
Note that the fourth equation in~\refb{ecombined} has already been
argued to hold with an equal sign.  We will not need to use 
the last equation
in~\refb{ecombined}, but will verify in section~\ref{sanalysis}  
that our solution~\refb{esolrecnewNN} for the $N$'s 
satisfies all the
equations in~\refb{ecombined} with an equal sign.

We will consider two situations below.  
First  will be the gluing of 
an open string three-point vertex to another vertex 
by an open string propagator.  This will allow us to fix
the sign in the third constraint in \refb{ecombined} when one of the surfaces
is the three open string vertex, with factor $N_{0,1,0,3}$.
Second, we will consider the effect of gluing a pair of disks, 
each with one closed string and at least one open string, via an open string
propagator. This allows us to fix the sign of the third constraint in \refb{ecombined} 
when both surfaces are disks with one closed and some open strings,
with factor $N_{0,1,1,n_{o1}} N_{0,1,1,n_{o2}}$ on one side and
$N_{0,1,2,n_{o1}+n_{o2}-2}$ on the other side.

\begin{enumerate}

\item[1.] 
Consider  
gluing a pair of open string three-point vertices to form 
a contribution to the 
four-point open string amplitude ${\cal A}_{0,1,0,4}$.
As described in P3, first bullet point,
the order in which we write the open string vertex 
operators in the correlator
 follows the order in which they appear as we traverse the boundary keeping
 the world-sheet  to the left.   
 We shall place the punctures
in the first interaction vertex at  $-\infty$, $-1$ and 0 and the punctures in the
second interaction vertex at 0, 1 and $\infty$ and use the punctures
at 0 to glue the
surfaces, choosing the local coordinates at the puncture at 0 to be 
$\lambda z$ for some constant  
$\lambda>1$ where $z$ is the upper
half plane coordinate itself. Then the gluing takes the form
\be 
\lambda^2 \, z_1z_2 =-q_o\, , 
\ee
and we can describe the 
sewn surface by the coordinate system $z_1$ in which the
punctures are located at $z_1=-\infty, -1, -q_o/\lambda^2, -q_o/\infty$. 
This situation is shown in Figure~\ref{FF4FF}. 
\begin{figure}[h]
	\centering
\epsfysize=6.2cm
\epsfbox{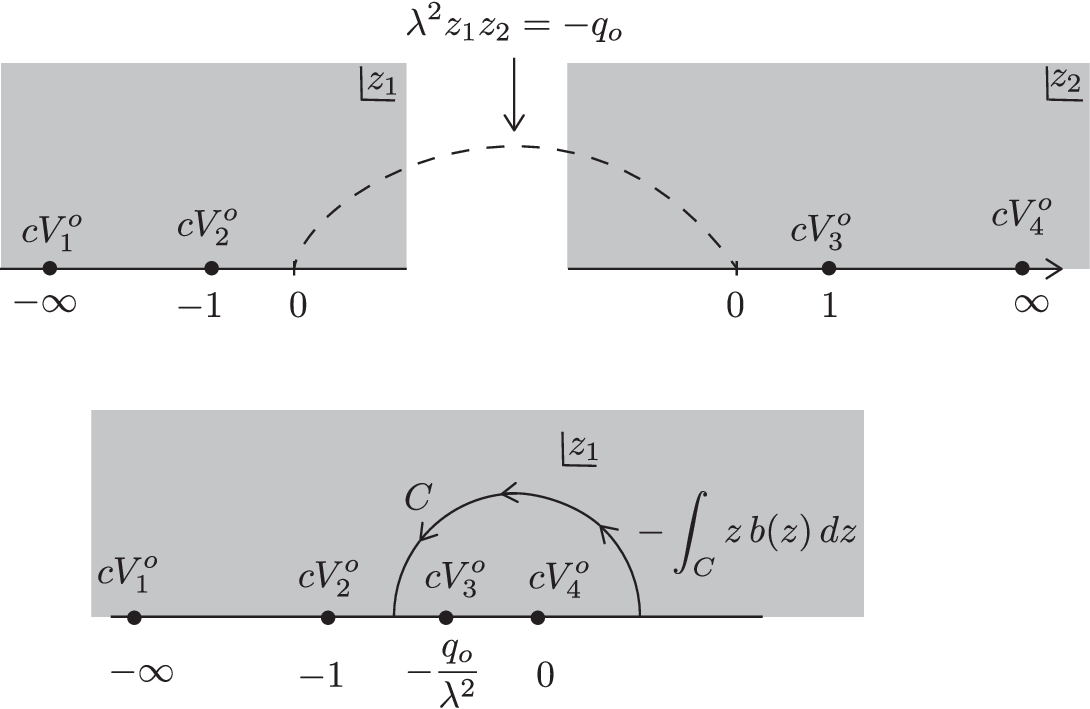}
	\caption{\small  
	The gluing of two three-punctured disks $z_1$ and $z_2$, represented as 
	upper-half plane regions,  via an open string propagator that implements
	gluing of the punctures at $z_1=0$ and $z_2=0$.  This results  
	in the  four-punctured disk, shown below, 
	including a $(-b_0)$ insertion, 
	and represented on the $z_1$ upper-half plane.} 
	\label{FF4FF}
\end{figure}

Using \refb{eopenpropX} we see that the resulting 
contribution to the four point amplitude takes the form
\ben 
&& {\cal A}_{0,1,0,4}^{\inc}  (cV^o_1, cV^o_2,cV^o_3, cV^o_4) \nonumber \\
&= & \  
 g_s N_{0,1,0,3} N_{0,1,0,3}\, \int_0^1 {dq_o\over q_o} \, 
\langle cV^o_1(-\infty) cV^o_2(-1) \, (-b_0)\,  cV^o_3(-q_o/\lambda^2) 
cV^o_4(0)\rangle\, .
\een
Expressing $b_0$ as $\ointop z b(z) dz$ with the contour enclosing the points
$-q_o/\lambda^2$ and 0 in the anti-clockwise direction, 
and deforming the $z$ contour to pick up residues from the poles
at $-q_o/\lambda^2$ and $0$, we get
\ben\label{efourpointcor}  
&& \hskip-30pt {\cal A}_{0,1,0,4}^{\inc}  (cV^o_1, cV^o_2,cV^o_3, cV^o_4)  \nonumber \\
&=  & 
  - g_s N_{0,1,0,3} N_{0,1,0,3}\, \int_0^1 {dq_o\over q_o} \, 
\langle cV^o_1(-\infty) cV^o_2(-1)  (-q_o \, \lambda^{-2}\, 
V^o_3(-q_o\, \lambda^{-2})) cV^o_4(0)\rangle\non\\
&=&  g_s N_{0,1,0,3} N_{0,1,0,3}\, \int_{-1/\lambda^2}^0 du \, 
\langle cV^o_1(-\infty) \, cV^o_2(-1)  \, V^o_3(u)\, cV^o_4(0)\rangle\, ,
\een
where $u=-q_o/\lambda^2$ denotes the location of the integrated vertex operator. 
The insertion
$\int_{-1/\lambda^2}^0 du V_3^o(u)$ 
is precisely the prescription P2
 for
 an integrated open string vertex operator in
$\Hom^{(0,1,0,4)}$.
Furthermore, we see that as desired, 
the unintegrated vertex operators at $-\infty$, $-1$ and 0
are placed inside the correlator in the anti-clockwise order. 
This shows that the  
correlator in \refb{efourpointcor} 
correctly represents $\Hom^{(0,1,0,4)}$ and hence the prefactor
can be identified as $g_s N_{0,1,0,4}$. This gives, without sign ambiguity, 
\be \label{enewoneab}
N_{0,1,0,4} =  N_{0,1,0,3} N_{0,1,0,3}\, .
\ee

\medskip 
\noindent

This is a special case of the third constraint in 
\refb{ecombined} with the $\sim$ replaced by $=$.
In fact one can easily
generalize 
the analysis above to get the more general relation
\be \label{enewonea}
N_{g,b,n_c,n_o+1} =  N_{g,b,n_c,n_o} N_{0,1,0,3}\, .
\ee
To prove this, we start from a Riemann surface $\Sigma_{g,b,n_c,n_o}$ and
identify an open string puncture where we have an unintegrated vertex operator
(possibly accompanied by a $\BB$ insertion to the right, following P2). 
We then glue this puncture to $\Sigma_{0,1,0,3}$ 
following the procedure described above. This will replace the unintegrated open 
string vertex operator by a pair of open string vertex operators, -- one integrated
representing the analog of $V^o_3(u)$ in \refb{efourpointcor} and the other unintegrated
representing the analog of $V^o_4(0)$ in \refb{efourpointcor}
(possibly accompanied by a $\BB$ to the right that was 
present in $\Sigma_{g,b,n_c,n_o}$).
This is in accordance with \refb{integopeninsert}, showing that we reproduce
$\Hom^{(g,b,n_c,n_o+1)}$ after the gluing.

\item[2.]
Next consider building a disk
amplitude  ${\cal A}_{0,1,2,0}$
of two closed strings in two different ways. The first way is to connect
a closed string three point function on sphere to a closed string one point function
on the disk by a closed string propagator. 
This is a special case of \refb{e433} but it will be useful to write down the explicit
expression for the correlation function. 
Let the closed string punctures on the sphere be located  
at $z_1$, $z_2$ and 0.
 We take the local coordinate at
0 to be the global coordinate $z$ on the complex plane and use it for gluing.
Using the result \refb{e425} and that
$N_{0,0,3,0}=1$, we can
express the result as
\ben\label{eb.13}  
{\cal A}_{0,1,2,0}^{\inc}  (c\bar c V^c_1, c\bar c V^c_2) &=& - g_s\, \AL\, 
N_{0,1,1,0}\,  \int_0^1 {d\Qr} \, \Qr^{-1}\, \Big\langle c\bar c V^c_1(z_1) c\bar c V^c_2(z_2) 
(e^{-\Lambda}\Qr)^{L_0+\bar L_0} b_0^+
\Big|B\Big\rangle\nonumber \\
&=& g_s \AL\, 
N_{0,1,1,0}\,  \int_0^1 {d\Qr} \, \Qr^{-1} \,
\Big\langle B\Big|  (e^{-\Lambda}\Qr)^{L_0+\bar L_0} b_0^+
c\bar c V^c_1(z_1^{-1}) c\bar c V^c_2(z_2^{-1}) 
\Big\rangle \nonumber \\
&=& - g_s \AL \, N_{0,1,1,0} \, \int_0^1 {d\Qr}  \,  
\Big\langle \BB\left[{\p\over \p \Qr}\right]  c\bar c V^c_1(
e^{-\Lambda} \Qr z_1^{-1}) c\bar c V^c_1(e^{-\Lambda} \Qr  z_2^{-1}) 
\Big\rangle\, , \nonumber \\
\een
where we have used $B[\p/\p \Qr]=-b_0^+/\Qr$ following the discussion 
above \refb{esewtoB}.
Comparing this with \refb{esewtoB} we see that 
the integrand in \refb{eb.13} can be identified as $-\Hom^{(0,1,2,0)}$ 
and hence the $\AL \, N_{0,1,1,0}$ factor 
can be identified as the 
normalization constant 
$N_{0,1,2,0}$ of the amplitude: 
\be\label{eb.23a}
N_{0,1,2,0} = \AL \, N_{0,1,1,0}\, .
\ee
This, as expected, is a special case of \refb{e433}.

The second way we can build a disk two-point function of two closed strings is to
take a pair of open-closed disk amplitudes and join them by an open string
propagator. The geometric interpretation of this is as follows. 
We take the open string punctures  
at $0$  and let $w_1=\lambda \, z_1$ and $w_2=\lambda\, z_2$ be the local 
coordinates at these punctures at the two disks,  
with $z_1$ and $z_2$ being the upper half plane global 
coordinates.
Let us place the closed string punctures at 
$z_1= i$ and $z_2 = i$.  
Under gluing, the coordinates $w_1$
and $w_2$ are identified via $w_1w_2=-q_o$. 
On the sewn surface, taken to be $z_1$  
upper half plane,  
 the closed string
punctures are at 
$z_1=i$ and $z_1=-q_o/(i\lambda^2) =iq_o/\lambda^2$ 
and the contribution to the 
amplitude takes the form:
\be\label{eb.15} 
{\cal A}_{0,1,2,0}^{\inc}  (c\bar c V^c_1, c\bar c V^c_2)  =  
- g_s\, N_{0,1,1,1}^2 \,\int_0^1 {dq_o\over q_o} \, \langle c\bar c V^c_1(i) \, b_0 \,
 c\bar c V^c_2(iq_o/\lambda^2)\rangle\, ,
\ee
with the contour defining $b_0$ enclosing the point $i q_o/\lambda^2$.
Using \refb{eminus} we can express \refb{eb.15} as,
\be\label{eb.15new}
{\cal A}_{0,1,2,0}^{\inc} (c\bar c V^c_1, c\bar c V^c_2) = g_s\, 
N_{0,1,1,1}^2 \,\int_0^1 {dq_o} \, \langle \BB\left[{\p\over \p q_o}\right]
c\bar c V^c_1(i) \, 
c\bar c V^c_2(i q_o/\lambda^2)\rangle\, . 
\ee
Our goal will be to turn the integrand
appearing here into the
integrand that appears to the right of the 
integral symbol
 in \refb{eb.13}. We can try to 
 manipulate \refb{eb.15new} 
to the form \refb{eb.13}, but we can take a short-cut by noting that in this case there
is no sign ambiguity involving where we place $\BB$, and both \refb{eb.13} and \refb{eb.15new}
are in accordance with the general prescription for ghost insertion 
given in \refb{e233open}. 
So the only thing we need to check to see if the integrands in \refb{eb.13} and \refb{eb.15new}
agree is to check if the orientation of the $\Qr$ integral in \refb{eb.13} agrees with the
orientation of the $q_o$ integral in \refb{eb.15new}. For this we note that if we decrease
$\Qr$ in \refb{eb.13} keeping the boundary fixed, then the two closed string vertex operators
come close to each other. On the other hand if we decrease $q_o$ in \refb{eb.15new} keeping
the boundary fixed, 
the distance between the pair of closed string punctures  
increase. Therefore decreasing
$\Qr$ in \refb{eb.13} will correspond to increasing $q_o$ in \refb{eb.15new}. 
The extra minus sign in \refb{eb.13} 
compensates for this mismatch and shows
that the orientations in the integrand
in \refb{eb.13} and \refb{eb.15new} match. Hence $ N_{0,1,1,1}^2$ can be identified
as $N_{0,1,2,0}$:
\be\label{eb.17}
N_{0,1,2,0} =  N_{0,1,1,1}^2\, .
\ee
This is a special case of the third constraint in
\refb{ecombined} but now the $\sim$ has been replaced by $=$.

We can consider a generalization of this analysis where the two open-closed
disk amplitudes are replaced respectively 
by  disk amplitudes with one closed and $n_{o1}+1$ 
open strings and disk amplitudes with one closed 
and $n_{o2}+1$ 
open strings.
In this case on each of the disks, one of the open string vertex operators will be 
inserted at a fixed point on the disk with 
unintegrated vertex operator and the other open string vertex operators 
will be integrated along the
boundary keeping the world-sheet to the left. We can now join them via 
an open string propagator, using the unintegrated open string vertex
operators  for gluing. This will generate a disk amplitude of
the same form as in \refb{eb.15new},  with $n_{o1}+n_{o2}$ 
additional open string insertions, 
all integrated along the boundary keeping
the world-sheet to the left. Since without the open string insertions the corresponding
amplitude gives correctly  the expression for 
$\Hom^{(0,1,2,0)}$, with the open string insertions it gives  
the expression for $\Hom^{(0,1,2,n_{o1}+n_{o2})}$. 
This gives the generalization of
\refb{eb.17}:
\be \label{eb.17gen}
N_{0,1,2,n_{o1}+n_{o2}}= N_{0,1,1,n_{o1}+1} N_{0,1,1,n_{o2}+1}\, .
\ee
\end{enumerate}

\sectiono{Analysis of the constraint equations and results for amplitudes} \label{sanalysis}

In this section we have several objectives.  Our first step will
be to find the solution of  
the constraint equations.   This will give
us a formula for the $N_{g,b,n_c,n_o}$ in terms of $\AL$. 
We then discuss various consequences of this formula.
Even though we will describe
 the analysis for bosonic string theory, 
the analysis for superstrings proceeds in the same way, leading to identical results.

\subsection{Solution of the constraint equations}

The analysis of section \ref{NormForm} has established four constraints that hold without sign
ambiguities.  These are \refb{eclosedseparating2}, \refb{eproptwo}, \refb{enewonea} and~\refb{eb.17gen}, collected here for convenience
\be\label{eclosedseparating2XX}
N_{g_1+g_2,b_1+b_2,n_{c1}+n_{c2}-2,n_{o1}+n_{o2}}
= \, \AL \, N_{g_1,b_1,n_{c1},n_{o1}} N_{g_2,b_2,n_{c2},n_{o2}}\, ,
\ee
\be \label{eproptwoXX}
N_{g+1,b,n_c-2,n_o} =\AL\, N_{g,b,n_c,n_o} \, , 
\ee
\be \label{enewoneaXX}
N_{g,b,n_c,n_o+1} =  N_{g,b,n_c,n_o} N_{0,1,0,3}\, ,
\ee
\be \label{eb.17genXX}
N_{0,1,2,n_{o1}+n_{o2}}= N_{0,1,1,n_{o1}+1} N_{0,1,1,n_{o2}+1}\, .
\ee
As special cases of these equations, we also have at our disposal 
\refb{e433}, \refb{eb.13a}, \refb{eb.23a} and \refb{eb.17}, also collected below
\be \label{e433XX}
N_{g,b+1,n_c,n_o} 
=\, \AL\, N_{g,b,n_c + 1,n_o} \, N_{0,1,1,0}\, ,  
\ee
\be \label{eb.13aXX}
N_{g,b+1,n_c-1, n_o+1} = \AL \, N_{0,1,1,1} \, N_{g,b,n_c,n_o}\, , 
\ee
\be\label{eb.23aXX}
N_{0,1,2,0} = \AL \, N_{0,1,1,0}\, , 
\ee
\be\label{eb.17XX}
N_{0,1,2,0} =  N_{0,1,1,1}^2\, .
\ee
Finally we have \refb{XYZ}
\be
\label{XYZnew}
 N_{g,b,n_c,n_o} \  =  \   N_{g,b-1,n_c,n_o+2} \, . 
 \ee
In this section we shall show 
that these results allow us to fix all the $N_{g,b,n_c,n_o}$.

Since $N_{0,0,3,0}$ has been set to one, 
one can then use \refb{eclosedseparating2XX}, \refb{eproptwoXX} to
show recursively that 
\be\label{ea33}
N_{g,0,n_c, 0} = \AL^{3g-3+n_c} \,. 
\ee
Now we add boundaries and open string punctures.
First consider adding {\em one}
boundary component with
$n'_o\geq 1 $ open string punctures.  For this we can first 
do closed string gluing of a surface of type 
$(0,1,1,1)$ to a Riemann surface
of type $(g,0,n_c+1,0)$ 
using \refb{eclosedseparating2XX}, so that we have
\be
N_{g, 1, n_c,  1} = \AL N_{0,1,1,1}  N_{g, 0, n_c+1, 0}
=\AL^2 \, N_{0,1,1,1}\, \AL^{3g-3+n_c} \, .
\ee
We can now use \refb{enewoneaXX}, that describes
the effect of gluing a 3-open string vertex, 
to add extra 
open strings, one by one, to get
\be 
N_{g, 1, n_c,  n'_o} =  (N_{0,1,0,3})^{n'_o-1}\, \AL^2\,  N_{0,1,1,1} \, 
 \AL^{3g-3+n_c} = \AL^2\,  {N_{0,1,1,1}\over N_{0,1,0,3}} \, 
 (N_{0,1,0,3})^{n'_o}\, 
 \AL^{3g-3+n_c} \, .
\ee
We can do this $\hat b$ times to add $\hat b$ boundaries that have
each at least one open string puncture.  Then we get
\be
\label{340oirejf}
N_{g, \hat b, n_c,  n_o} =   \AL^{2\hat b}  \Bigl({N_{0,1,1,1}\over N_{0,1,0,3} } 
\Bigr)^{\hat b} (N_{0,1,0,3})
^{n_o}\ \AL^{3g-3+n_c}
\ee
with $n_o$ the total number of open string punctures. 

Now consider adding boundaries without punctures. Using \refb{e433XX} 
we see that
each time we do this we have an extra factor $\gamma$
with 
\be
\gamma = \AL \cdot \AL\,  N_{0,1,1,0} = \AL\,  N_{0,1,1,1}^2 \,,
\ee
where the first factor of $\AL$ in the second expression comes 
from expressing $N_{g,b,n_c + 1,n_o}$ on the right hand side of 
\refb{e433XX} as $\AL\, N_{g,b,n_c,n_o}$ using
\refb{340oirejf} 
and in the last step
we used the relations \refb{eb.23aXX} and  \refb{eb.17XX}. 
To add $b_0$ boundaries we need a factor $\gamma^{b_0}$ in~\refb{340oirejf}
and we get
\be
\label{340oirejfxx}
N_{g,\,  b= \hat b + b_0, \, n_c,  n_o} =   \AL^{2\hat b}  \Bigl({N_{0,1,1,1}\over N_{0,1,0,3} } 
\Bigr)^{\hat b} (N_{0,1,0,3})
^{n_o}\  (\AL N_{0,1,1,1}^2)^{b_0}   \AL^{3g-3+n_c} \, . \ 
\ee
Using this in \refb{eb.17genXX} with $n_{o1}+n_{o2}\ne 0$ so that we have $\hat b=1, b_0=0$
on the left hand side, we get
\be
\AL^{2}  \Bigl({N_{0,1,1,1}\over N_{0,1,0,3} } 
\Bigr) (N_{0,1,0,3})^{n_{o1}+n_{02}} \AL^{-1}  
= \AL^{4}  \Bigl({N_{0,1,1,1}\over N_{0,1,0,3} } 
\Bigr)^2 (N_{0,1,0,3})^{n_{o1}+n_{02}+2}  \AL^{-4} \, ,
\ee
leading to,
\be
N_{0,1,1,1}  =\,  {\AL \over N_{0,1,0,3}}\, .
\ee
Substituting this into \refb{340oirejfxx} we see that the result depends only on the
total number of boundaries $b=\hat b + b_0$:
\be
\label{340oirejfxxxsxs}
N_{g, b, n_c,  n_o} =    (N_{0,1,0,3})^{n_o - 2b}   \AL^{3g-3+3b+n_c}\, . 
\ee
Finally, using \refb{XYZnew} for $(g,b,n_c,n_o)= (0,2,0,1)$ 
we get 
\be \label{enuexpression} 
 (N_{0,1,0,3})^{-3} \AL^3 =  N_{0,1,0,3}  \quad \to \quad 
  N_{0,1,0,3} =  \AL^{3\over 4} \,,
 \ee
and hence \refb{340oirejfxxxsxs} takes the form:
\be
\label{esolrecnewNN}
N_{g, b, n_c,  n_o} 
=   \  \AL^{3g-3+n_c + {3\over 2} b + {3\over 4} n_o}\,.
\ee
It is easy to check that \refb{esolrecnewNN} satisfies all the equations in 
\refb{ecombined} with $\sim$ replaced by $=$.

Equation~\refb{esolrecnewNN} 
is the unique solution of the constraint equations. 
This is the result quoted  
in~\refb{esolrecnewNNX}.  These $N$'s 
accompany the canonical forms $\Hom$ that are defined with 
{\em standard}
correlators of the original (un-normalized) 
string fields, with all insertions using prescriptions P1, P2, and P3.

\subsection{$K, g_o, \TT$, and amplitudes}  

\medskip
\noindent
{\bf Open string coupling:}  
We now relate the open string coupling $g_o$ 
to the constant $K$ 
and the string coupling $g_s$ of the open-closed
theory.  For this it suffices to examine the tree-level 
quadratic and cubic terms for the open string field.  These take the form
\be
\label{ocsftto2+3} 
\begin{split}
S_{\rm open}^{(2+3)} (\Psi_o) =\ & 
\tfrac{1}{2} \langle \Psi_o| Q_o |\Psi_o\rangle + 
\tfrac{1}{3!} \   g_s^{1/ 2} \,  N_{0,1,0,3} \,  \Hom^{(0,1,0,3)}  (\Psi_o , \Psi_o , \Psi_o)   
 \\[1.0ex]
= \ &  \tfrac{1}{2}  K \,
 \langle \Psi_o| Q_o| \Psi_o\rangle' 
+  
\tfrac{1}{3!}  \ g_s^{1/ 2} \,  N_{0,1,0,3} \, K  \,
\Hom^{\prime (0,1,0,3)}  (\Psi_o , \Psi_o , \Psi_o)  
\\[0.6ex] 
= \ &  \tfrac{1}{2} \, \langle \psi_o| Q_o| \psi_o\rangle' +  
\tfrac{1}{3!}  \ g_s^{1/ 2} \, \AL^{3/ 4}\, K^{-{1\over 2} }\,
\Hom^{\prime (0,1,0,3)}  (\psi_o , \psi_o , \psi_o)\, , \\
\end{split}
\ee
where in the last step we introduced canonically 
normalized open string fields $\psi_o\equiv K^{1/2}\Psi_o$ 
and recalled the value of $N_{0,1,0,3}$ from~\refb{enuexpression}.
We can now compare the above expression with the standard 
expansion of tree-level open string field  
in which the open string coupling constant $g_o$
appears in the cubic term 
\be
S_{\rm open}^{(2+3)} = 
\tfrac{1}{2}  \langle \psi_o, Q_o\, \psi_o\rangle' +  
\tfrac{1}{3!}  \, g_o\, \Hom^{\prime(0,1,0,3)}( 
\psi_o , \psi_o  ,\psi_o )  \, .
\ee
Comparing the cubic term here
with the cubic term in~\refb{ocsftto2+3} we read 
\be\label{edefgo}
g_o = 
  g_s^{1/ 2} \, K^{-1/2} \, \AL^{3/4}\, .
\ee
Squaring we get 
\be
\label{gopgopc}
g_o^2 =  g_s  \  K^{-1} \, \AL^{3/2} \,. 
\ee
This is the relation between the open and closed string field theory couplings. 
This relation is not completely fixed until we get the value of $K$.

\medskip
\noindent
{\bf Rewriting of the open-closed string amplitudes:}  
Having determined
the normalization
coefficients $N_{g,b,n_c,n_o}$ in~\refb{esolrecnewNN}, 
the string theory amplitudes  
corresponding
to a given value of $(g,b,n_c,n_o)$ may be expressed as
\ben\label{etermone}   
&& \hskip-30pt\AAA_{g,b,n_c,n_o}( A^c_1,\cdots, A^c_{n_c}; A^o_1,\cdots, A^o_{n_o} 
) \nonumber \\[1.5ex]
&=& 
g_s^{-\chi_{g,b,n_c,n_o}} \,  
\AL^{3g-3+n_c+{3\over 2} b+{3\over 4} n_o} 
\int_{\MM_{g,b,n_c,n_o}}
 \hskip-30pt \Hom^{(g,b,n_c,n_o)}(A^c_1,\cdots, A^c_{n_c}; A^o_1,\cdots, A^o_{n_o} ) \, .
\een
Using  $A^c_i=c\bar c V^c_i$, $A^o_i=K^{-1/2} c W^o_i$ 
and $\Hom'=K^{-b} \Hom$
as given in
\refb{edefacao}, \refb{edefW} and \refb{edefHomp}, we can express this as
\ben\label{etermtwonew}
&&\hskip-30pt \AAA_{g,b,n_c,n_o}(A^c_1,\cdots, A^c_{n_c}; A^o_1,\cdots, A^o_{n_o} 
)  \nonumber \\
&=& 
\,   g_s^{-\chi_{g,b,n_c,n_o}} \,
 N'_{g,b,n_c, n_o}  
\int_{\MM_{g,b,n_c,n_o}} \hskip-30pt \Hom^{\prime(g,b,n_c,n_o)}(c\bar c V^c_1,\cdots, c\bar c V^c_{n_c}; 
c W^o_1,\cdots, cW^o_{n_o} ) \, , 
\een
where the $N'$ coefficients are 
\be \label{edefnprime}
 N'_{g,b,n_c, n_o} \equiv  K^{-{1\over 2} (n_o-2b)} \AL^{3g-3+n_c
 +{3\over 2} b+{3\over 4} n_o}  \,. 
\ee

When applying the above expressions for the disk one-point function
of a closed string state, 
one must recall that on account of~\refb{e14} 
this amplitude
carries
an extra factor of $\AL$.  
We thus find
\be \label{etensionform} 
\AAA_{0,1,1,0}( c\bar c V^c) = 
\AL^{-{1\over 2}} \,  
\Hom^{ 
(0,1,1,0)}(c\bar c V^c) = \AL^{1/2} \, \langle c_0^- c\bar c V^c\rangle
 = \AL^{1/2} K  \langle c_0^- c\bar c V^c\rangle' \,.
 \ee

\medskip
\noindent
{\bf D-brane tension:} 
 We shall now  find the relation between the
constant $K$ and the tension $\TT$ of the corresponding D-brane,
assumed to be a D$p$-brane.   
For this
we consider a closed
string field of the form 
\be \label{e1pp4rep}
{\cal O}_h \,  =  c\bar c V_h = 
 \,{1\over g_s } h_{\mu\nu} c\bar c \p X^\mu \bar \p X^\nu\, ,
\ee
with constant $h_{\mu\nu}$.
Using~\refb{etensionform} we have the 
one-point function of the  closed string field
\be 
\{ {\cal O}_h \}_{0,1}
= \AL^{1/2} \, 
{1\over g_s } \,h_{\mu\nu} 
\langle c_0^- c\bar c \p X^\mu \bar \p X^\nu\rangle 
\, . 
\ee
We can evaluate the correlator using the doubling trick that replaces
$\bar \p X^\nu(z)$ by
$\p X^\nu (z^*)$ 
for directions tangential to the brane and  by 
$-\p X^\mu(z^*) $ for directions transverse to the
brane
and $\bar c(z)$ by $c(z^*)$ 
where $z^*$ is the complex conjugate of $z$.  
We also have the
operator product expansion 
$\p X^\mu(z) \p X^\nu(w) = -\eta^{\mu\nu} / 2(z-w)^2$  
and
the normalization
condition \refb{eopencorr}.
This gives 
\be \label{esecondcomp}  
\{ {\cal O}_h \}_{0,1}
= \AL^{1/2} K {1\over g_s } \, (h^\parallel_{\mu\nu}  - h^\perp_{\mu\nu})\  {1\over 2} \, \eta^{\mu\nu}\,  (2\pi)^{p+1}
\delta^{(p+1)}(0)\, ,
\ee  
where $h^\parallel_{\mu\nu}$ and $h^\perp_{\mu\nu}$ denote components of
$h_{\mu\nu}$ along the brane and 
transverse to the brane respectively. 
This is the contribution to the string field theory action to linear order in $h_{\mu\nu}$.

On the other hand we have shown in appendix \ref{sa} that the background
\refb{e1pp4rep} corresponds to switching on a background metric and dilaton 
as follows:
\be\label{ephiexp}
g_{\mu\nu} = \eta_{\mu\nu} + h_{\mu\nu}, \qquad
\phi = \tfrac{1}{4} \, \eta^{\mu\nu} h_{\mu\nu}\, .
\ee
Consider now the action $S_p$ of a D$p$-brane in this background:  
\be
S_p \ = \ -\TT\, \int d^{p+1}x \,  
e^{-\phi} \sqrt{-g }\, .
\ee
Using \refb{ephiexp} we can express 
the terms linear in $h_{\mu\nu}$ in this expression as 
\be\label{efirstcomp}
S_p\bigl|_h =  -\, {\TT} \, \left({1\over 2} \eta^{\mu\nu} \, 
h^\parallel_{\mu\nu}(x_\perp=0)  -  \phi(x_\perp=0)\right)
\int d^{p+1}x = -\tfrac{\TT}{4}\, \eta^{\mu\nu}\, (h^\parallel_{\mu\nu} 
- h^\perp_{\mu\nu}) (2\pi)^{p+1}
\delta^{(p+1)}(0)\, .
\ee
Comparing \refb{efirstcomp} with \refb{esecondcomp} 
we get
\be\label{eKTrelx}
K  
= -g_s    {\TT\over 2\sqrt \AL}\, .
\ee
We can now use this result  
and \refb{gopgopc} to get the anticipated coupling constant/tension relation~\refb{e3100int}:
\be\label{e3100}
g_o^2 =   
 - {2\over \TT} \AL^2   
= {1\over 2\pi^2 \TT}\,  \quad \to \quad  \TT\, g_o^2 =   {1\over 2\pi^2} \,.
\ee

Using \refb{edefgo} and \refb{eKTrelx} we can express 
\refb{etermtwonew}, \refb{edefnprime} 
in terms of $g_s$, $g_o$, and $\TT$,    
\be
\begin{split}
\label{efres}
\AAA_{g,b,n_c,n_o}( A^c_1,\cdots, A^c_{n_c}; A^o_1,\cdots, A^o_{n_o} )  
=\ &  \ g_s^{2g-2+2b+n_c} \, g_o^{n_o} \, \left(-\tfrac{\TT}{2}\right)^{b}\,  \AL^{3g-3+n_c+b} 
\\[0.8ex]
&  
\times 
\int_{\MM_{g,b,n_c,n_o}} \hskip-30pt
 \Hom^{\prime (g,b,n_c,n_o)}(c\bar c V^c_1,\cdots,c\bar c V^c_{n_c};
c W^o_1,\cdots, c W^o_{n_o}) \, . 
\end{split}
\ee
We will make repeated use of this form of the amplitude. 

An alternative expression  
for the amplitudes is given by 
\be  
\label{e549newerXX} 
\begin{split}
  \hskip-10pt\AAA_{g,b,n_c,n_o}(A^c_1,\cdots, A^c_{n_c};A^o_1,\cdots, A^o_{n_o}) 
 = \,  \  &  
 g_s^{-\chi_{g,b,n_c,0}} \ g_o^{n_o} K^b \  
 \AL^{{1\over 2} {\rm dim}_R \MM_{g,b,n_c, 0}} 
     \\[0.5ex]
  & \hskip-10pt  \times 
  \int_{\MM_{g,b,n_c,n_o}} \hskip-30pt\Hom^{\prime (g,b,n_c,n_o)} 
(c\bar c V^c_1,\cdots, c\bar c V^c_{n_c}; c W^o_1,\cdots, c W^o_{n_o})  \, .
\end{split} 
\ee
In this form, the powers of closed string coupling and the closed string factor $\eta_c$ are
controlled by the Euler number and real dimension of moduli space for
surfaces with zero open string punctures.  The power of $K$ is controlled by the
number of boundaries and  the power of $g_o$ is controlled by the number of external
open strings. 
Note, however, that  since $K,g_s,$ and $g_o$ are related by 
\refb{gopgopc},   there are many other ways of writing this formula.

\medskip
\noindent
{\bf Determination of $K$:}  
This evaluation is needed to 
compute $\TT$
and $g_o$ via  \refb{eKTrelx} 
and \refb{edefgo}.
As we discussed earlier, the determination of $K$ requires the consideration
of an amplitude where an open string propagator joins two open strings that lie
on the same boundary.  Thus, the simplest object that determines $K$ is the
annulus one point function of an external open 
string. 
This can be constructed in two 
ways. The first one is 
the defining construction using the closed string channel, where 
we use \refb{ebinsertX}, 
\refb{eomegadef} with $(g,b,n_c,n_o)=(0,1,0,1)$. 
The other construction takes an open string three point function on the disk and
joins two of the open strings by an open string propagator. 
This is the gluing condition given in 
\refb{4thstrictlyX} 
for $(g,b,n_c,n_o)=(0,2,0,1)$.  
This will have a modular
parameter $q_o$ associated with the open string propagator. 
Once we express the open string modulus $q_o$ in terms of the closed string modulus
$\Qr$
appearing in \refb{ebinsertX}, we can compare the two expressions for
$\Hom^{(0,2,0,1)}$  
obtained from these two different approaches. However, now there
will be a net excess power of $K^2$ in the closed string channel expression
and demanding equality of the two sides, we determine $K$.
This in turn will determine $\TT$ using \refb{eKTrel}.
The actual expression for $K$ (and hence for $\TT$) will depend on the details of the
open string spectrum and hence will be different for different D-branes.

A simpler approach to computing the tension (or equivalently $K$) is 
to compare the
annulus zero point function in the closed and 
open string channels\cite{Polchinski:1998rq,Polchinski:1998rr}. Even though the 
zero-point functions are not usually included as part of the string field theory 
action,\footnote{Zero-point functions contribute constants to the Batalin-Vilkovisky
SFT action, so they do not affect the verification of the master equation. Still, they are relevant
for background independence, and the associated puncture-free Riemann surfaces feature
in the geometric master equation.}  
we shall review the analysis here due to its simplicity.
For this let us consider the sum of the kinetic term of the closed string field
in the Siegel gauge and the disk one-point function of the closed string 
field:\footnote{We have not included the $e^{-\Lambda L_+}$ 
term accompanying $|B\rangle$
since this can
be absorbed into a shift in the integration variable $s$ in \refb{eTtwo}.}
\be\label{eTone}
S^c_{\rm kin} + S_{\rm 1pt}^c=  
\tfrac{1}{2}  \langle \Psi_c|c_0^- c_0^+ L_0^+ |\Psi_c\rangle 
+\eta_c^{1/2} \langle \Psi_c | c_0^- |B\rangle \, .
\ee
Eliminating $|\Psi_c\rangle$ by its equation of motion gives 
$|\Psi_c\rangle
= - \eta_c^{1/2}\, b_0^+ (L_0^+)^{-1} |B\rangle$ and substituting it back into the
action we get the value of the annulus zero-point function:
\be\label{eTtwo}
{\cal A}_{\rm ann} =  
-\tfrac{1}{2}  \, \eta_c \, \langle B| c_0^- b_0^+ (L_0^+)^{-1} |B\rangle=
-\tfrac{1}{2}  \, \eta_c \,  \int ds\,  \langle B| c_0^- b_0^+ \, e^{-s L_0^+} |B\rangle\, .
\ee
Here $s$ has the interpretation as the height of the cylinder, with the circumference
set to $2\pi$.

This must coincide 
with the annulus 
partition function ${\cal A}_{\rm ann}$ 
computed in the 
open string channel, given by the logarithm of the 
one loop determinant of the open string fields in the Siegel gauge:
\be \label{eTthree}
{\cal A}_{\rm ann} = \ln\Biggl[\prod_b h_b^{-1/2} \prod_f h_f^{1/2}\Biggr] = \int {dt\over 2t} 
\Biggl (\sum_b e^{-t h_b}- \sum_f e^{-t h_f}\Biggr) = \int {dt\over 2t} \, {\rm Str}[e^{-t L_0}]\, ,
\ee
where $h_b$ and $h_f$ are the $L_0$ eigenvalues of the bosonic and fermionic open
string states in the Siegel gauge and 
Str denotes trace over Siegel gauge states, counting bosonic states (carrying
odd ghost number vertex operators) with plus sign and fermionic states with minus
sign.  Here $t$ has the interpretation as the circumference of the cylinder whose height
is fixed at $\pi$.

The parameters $s$ and $t$ can be related to each other by comparing the ratio of
the height to the circumference in each description. This gives
\be \label{eTfour}
t/\pi = 2\pi / s\, .
\ee
With the help of this relation we can compare the integrands in
\refb{eTtwo} and \refb{eTthree}. Since
\refb{eTtwo} is proportional to $K^2$,  each boundary state being proportional
to $K$, and \refb{eTthree} has no explicit  factor of $K$, this 
comparison
determines $K$. 
We have illustrated
the use of this procedure in appendix \ref{sb} by computing the
tension of D$p$-brane in the bosonic string theory 
in twenty six dimensional Minkowski space-time. 

\sectiono{Application to on-shell amplitudes}  \label{samplitudes}

While \refb{efres} gives the normalization of the 
amplitudes, we must carefully follow the sign
conventions described earlier for defining the 
$\int\langle\cdots \rangle'$ 
that
appears in \refb{efres}. 
We shall illustrate this by describing some specific application of our results
for the usual on-shell amplitudes in string theory.

\begin{enumerate}
\item First consider the closed string one point function on the disk.
Using \refb{etensionform} and \refb{eKTrelx}, we get  
\be \label{ediskone}
\AAA_{0,1,1, 0} (c\bar c V^c) = -g_s\, \tfrac{\TT}{2} 
\langle c_0^- c\bar c \, V^c\rangle' \, .
\ee
\item
Next, let us 
consider the 
disk amplitude with one closed string puncture and one open string puncture.
With closed string vertex operator 
$A^c=c\bar c V^c$
and an open string vertex operator $A^o=K^{-1/2} c W^o$,  
we have, from \refb{efres} 
\be\label{e463a} 
\begin{split}     
\AAA_{0,1,1,1}(A^c;A^o)  
=   -
i\pi\TT\, g_s\, g_o\, \langle c\bar c V^c \, c W^o\rangle'\, .
\end{split}
\ee
In arriving at this equation we need to take into account the 
minus sign in \refb{esignchange} which tells us that
$\Hom^{\prime(0,1,1,1)}(c\bar c V^c, c W^o)$
needs to be interpreted as 
$-\langle c\bar c V^c \, c W^o\rangle'$. 
Any further
addition of open string states to this amplitude will follow the
P2 rule
and as a result, all of these amplitudes will also 
carry an extra minus sign. 
\item 
Let us suppose that we have an amplitude 
with fixed set of external open and closed string states, given by
an appropriate integral over the moduli space of an appropriate Riemann surface, 
and we want to
study the new amplitude  obtained 
by inserting an additional closed string   
state $A^c=c\bar c V^c$ 
into the original amplitude.
It follows from 
\refb{efres} 
and \refb{e243} that the presence of the extra state corresponds 
to inserting into the properly normalized world-sheet 
correlation function for the original amplitude
 an extra integrated operator  
\be \label{e243rep}
- g_s \, \AL\, 
\int dy \wedge d\bar y \,  
V^c (z=y)    
 = -{g_s\over \pi} \int dy_R dy_I \, 
 V^c(y)  \,.             
\ee  
Here we used $y=y_R+iy_I$ 
and that $dy_R\wedge dy_I$ gives the positive 
integration measure $dy_Rdy_I$. 

\item We can give the analogous result for the insertion of open 
string state $A^o = K^{-1/2} cW^o$ into an open-closed amplitude. 
Using 
\refb{efres} and \refb{integopeninsert}, we see that 
for every insertion of an open string state 
 into an
amplitude for a dimension one matter primary $W^o$  
we insert into the world-sheet
correlator the integrated operator
\be
g_o \int dx W^o(x) \, .
\ee

\item From \refb{efres}, \refb{esewtoB} 
and the comments below \refb{esewtoB}, we see 
that the disk two-point function of a pair of closed string states with vertex operators
$A^c_1=c\bar c V^c_1$, $A^c_2=c\bar c V^c_2$ 
has the general structure 
\be\label{exx1}
\AAA_{0,1,2,0}(c\bar c V^c_1, c\bar c V^c_2)=
- g_s^2 \, 
\  \tfrac{\TT}{2}\, \int \Hom^{\prime (0,1,2,0)} (c\bar c V^c_1, c\bar c V^c_2)
= - g_s^2 \, \tfrac {\TT}{2}\, \int 
dq \, \langle \BBB[{\p/ \p q}] \, 
c\bar c V^c_1 \, c\bar c V^c_2\rangle'\, ,
\ee
where the modular parameter $q$ has to be chosen so that 
{\it as $q$ increases, the  
distance between the two vertex operators decreases}. 
We can work in the upper half plane, place $c\bar c V^c_1$ at a fixed point $i$ and
$c\bar c V^c_2$ at a variable point $iy$. Denoting by $z$ the upper half plane coordinate
and taking $w=z - iy$ as the local coordinate at the second puncture, we get 
\be\label{eBB5}
\BB\left[{\p\over \p y}\right]
= -i\bigg[\ointop_{iy} b(z) dz - \ointop_{-iy} \bar b(\bar z) d\bar z\bigg]\, , 
\ee
with the minus sign having the same origin as in \refb{ebboneopenxx}. 
Furthermore 
the
direction of increasing $y$ gives positive integration measure since in this direction the
two closed string punctures approach each other.
Inserting \refb{eBB5} into \refb{exx1}, we get,
\be
\AAA_{0,1,2,0}(c\bar c V^c_1, c\bar c V^c_2)=
{i\over 2}\,  g_s^2 \, \TT\,  \int dy\,  \left\langle c\bar c V^c_1(i)\,  
(c+\bar c) V^c_2(iy) \right\rangle'  
\, .
\ee
This can also be derived from \refb{eb.15} after 
using \refb{esolrecnewNN} and \refb{eKTrelx}.
\end{enumerate}

Finally, note that some of the 
signs mentioned above can be changed by field redefinition,
{\it e.g.} a redefinition of the form $\Psi_c\to -\Psi_c$ will lead to an extra factor of $(-1)^{n_c}$
in the amplitude and change the minus sign to plus sign on the right hand sides of
\refb{ediskone},  
\refb{e463a} and 
\refb{e243rep}.\footnote{This is the convention 
used in \cite{Eniceicu:2022xvk,Alexandrov:2023fvb}.} 
A similar transformation can be done for the open string fields. 
However once a few signs have been chosen this way, the
signs of all other amplitudes get fixed according to the analysis given above.

\bigskip

\noindent{\bf Acknowledgement:} 
The work of A.S. was supported by ICTS-Infosys Madhava 
Chair Professorship, the J. C. Bose fellowship of the Department of Science
and Technology, India and the Department of Atomic Energy, Government of India, under project no. RTI4001. This research was also 
supported in part by grant NSF PHY-2309135 to the 
Kavli Institute for Theoretical Physics (KITP).

The work of B.Z was supported by the U.S. Department of Energy, Office of Science, Office of High Energy Physics of U.S. Department of Energy under grant Contract Number  DE-SC0012567 
(High Energy Theory research). 

\appendix

\sectiono{From SFT 
background to EFT 
background} \label{sa}

Our goal in this appendix will be to relate the string field theory 
(SFT) background
 deformation given in
\refb{e1pp4rep} to the deformation of the metric and dilaton of the effective field theory (EFT),
eventually establishing \refb{ephiexp}.

We first note that 
in the presence of a background target space string metric $g_{\mu\nu}$,
the world-sheet action $S_{\rm ws}$ 
associated with the non-compact coordinates takes the form
\be
S_{\rm ws} = -{1\over 4\pi } \int dx dy \, g_{\mu\nu} (\p_{x} X^\mu \p_{x} X^\nu 
+ \p_{y} X^\mu \p_{y} X^\nu ) = 
-{1\over \pi   } \int dx dy  \, g_{\mu\nu} \p X^\mu \bar \p X^\nu, \quad
z\equiv x + i y\, ,
\ee
with $x, y$ real coordinates on the world-sheet, and $\partial$ and $\bar \partial$
derivatives with respect to $z$ and $\bar z$, respectively.
Writing 
\be
g_{\mu\nu}=\eta_{\mu\nu}+h_{\mu\nu}\, ,
\ee 
we see that to first order in $h_{\mu\nu}$, the
deformation by $h_{\mu\nu}$ corresponds to the insertion of the operator
\be\label{eapp2}
S_{\rm ws}|_{h} \ = \ -{1\over \pi  } \int dx dy
 \ h_{\mu\nu} \,  \p X^\mu \bar \p X^\nu\, ,
\ee
in the world-sheet correlator. On the other hand, it follows from
\refb{e243rep} 
that if we deform the background string field by 
$c\bar c V$ 
for some
dimension (1,1) matter primary $V$, it corresponds to inserting into the world-sheet
correlator a term
\be\label{eapp3}
- g_s\, {1\over \pi} \int dx dy  
\, V\, .
\ee
Comparing 
the last two equations  
we can read $V$ and 
see that in order to turn on a metric 
deformation $h_{\mu\nu}$, we need to turn on a string field background 
${\cal O}_h = c\bar cV$ as given in \refb{e1pp4rep}:
\be \label{e1pp4}
{\cal O}_h  = {1\over g_s   } h_{\mu\nu}\   c\bar c \p X^\mu \bar \p X^\nu\, 
= {1\over g_s} ( - \tfrac{1}{2}  h_{\mu\nu} ) \, c\bar c \ i\sqrt{2} \p X^\mu \
i\sqrt{2} \bar \p X^\nu\, . \ee
However this could also turn on a dilaton field. Our goal now will be to find the value of
the dilaton field that is turned on.
For this we follow~\cite{Yang:2005rx}.

Let us expand the
closed string field associated with the massless 
fields as
\be \label{epsiexpand}
\begin{split}
\ket{\Psi_c} & =  {1\over g_s } 
\int {d^Dk\over (2\pi)^D} ~ \Bigl[  - \tfrac{1}{2}
h_{\mu\nu} ( k) \,\alpha_{-1}^\mu \bar \alpha_{-1}^\nu \, c_1 \bar c_1
 + e(k)  \, (c_1 c_{-1}    - 
\, \bar c_1 \bar c_{-1}) \\[0.2ex]
&\hskip70pt+ i\sqrt{\tfrac{1}{2} }\,
\,  f_\mu (k) \, c_0^+ \bigl(c_1 \alpha_{-1}^\mu 
-  \bar c_1 \bar \alpha_{-1}^\mu\bigr) \, \Bigr] \ket{k} \,. 
\end{split}
\ee
Note that we have restricted the expansion to a linear combination of states each of which
is anti-symmetric under the exchange $c_n\leftrightarrow \bar c_n$, 
$\alpha^\mu_n\leftrightarrow \bar\alpha^\mu_n$. 
As a result the two form field is not included in this expansion. 
Recalling that $\alpha^\mu_{-n}$ are the oscillators of $i\sqrt {2}\p X^\mu$, and similarly for the barred oscillators,
we can identify $h_{\mu\nu}$ appearing in \refb{e1pp4} with the one appearing in
\refb{epsiexpand}. In order to represent the background \refb{e1pp4} we need to
set $f_\mu$ and $e$ to zero in \refb{epsiexpand}, but for now we proceed by keeping
them.

The closed string field given in \refb{epsiexpand} is annihilated by
$b_0^-$ and $L_0^-$ as required but does not satisfy the Siegel gauge condition.
For closed strings BPZ conjugation can be implemented with the map $z \to 1/z$
and the result is that for the oscillators
of a dimension $d$ field we have~\cite{Zwiebach:1992ie} 
\be
\hbox{bpz} (\phi_n) = (-1)^d \phi_{-n} \,,
\ee
all oscillators transforming with the same sign prefactor, independent 
of the mode number.  It then follows that
 BPZ conjugate of the above string field, needed for the construction of
the kinetic term in the action, is given by
\be \label{eexpansionclosed}
\begin{split}
\bra{\Psi_c} & = 
{1\over g_s } \int {d^D k\over (2\pi)^D} ~ \bra{k}\Bigl[  - \tfrac{1}{2}
h_{\mu\nu} (k) \,\alpha_{1}^\mu \bar \alpha_{1}^\nu \, c_{-1} \bar c_{-1}
 + e(k)  \, \bigl( c_{-1} c_{1}    - \bar c_{-1} \bar c_{1}\bigr) \\
&\hskip85pt- i\sqrt{\tfrac{1}{2} }
\,  f_\mu (k) \, c_0^+ \,\bigl( c_{-1} \alpha_{1}^\mu 
- \bar c_{-1} \bar \alpha_{1}^\mu\bigr)\Bigr] \,.
\end{split}
\ee

We now 
construct
the quadratic term of the gauge invariant
bosonic string action~(\ref{esftactiongaugefixed}), given by
\be
\label{quad-action-def-89}
S^{(2)} =   \tfrac{1}{2}  \bra{\Psi_c} \, c_0^-
Q_c \ket{\Psi_c} ~ \,.
\ee
Here $Q_c=Q_B+\bar Q_B$ is the (ghost-number one) BRST operator of the 
conformal field theory. 
The level expansion of the BRST operator 
gives:
\be
\begin{split}
Q_c = & \ \  \,c_0^+\Bigl(  \tfrac{1}{2} k^2 - 2 \Bigr)  \\
&   + c_0^+\Bigl(  \alpha_{-1}\cdot \alpha_1 + b_{-1} c_1 + c_{-1} b_1  
\ + \  \bar\alpha_{-1}\cdot \bar\alpha_1 + \bar b_{-1} \bar c_1 + \bar c_{-1} \bar b_1 \Bigr)  \\
 & \ \  + \sqrt{\tfrac{1}{2}} \,  
  k  \cdot \big( \alpha_{-1}  c_1 + c_{-1} \alpha_1 \bigr)
+ \sqrt{\tfrac{1}{2}} \, k\cdot \big( \bar\alpha_{-1} \bar c_1 + 
\bar c_{-1} \bar\alpha_1 \bigr)\\[0.5ex]
 & \ \ - b_0^+ ( c_{-1} c_1 + \bar c_{-1} \bar c_1 ) + \ldots\, \, ,
\end{split}
\ee
where we have dropped terms proportional to $L_0^-$
and $b_0^-$.  
Using (\ref{sl2c_overlap}) we now get,
\be
\label{quad-action}
\begin{split}
\,S^{(2)} &=  \frac{1}{8  g_s^2}\, \int d^Dx \, \,    
\Bigl[\, \, \tfrac{1}{4}  h^{\mu\nu} \, \Box \, h_{\mu\nu} 
\,-\, 2\, {e} \, \Box \, e
- 2\, f^\mu\, f_\mu  - 2\, f^\nu  \,
\bigl(\, \p^\mu h_{\mu\nu} \, + 2\, \p_\nu e \bigr)\Bigr]\,.
\end{split}
\ee

The quadratic string action (\ref{quad-action-def-89}) is invariant under the gauge transformations
\be 
\label{gtfa}
\delta \ket{\Psi_c}=Q_c  \ket{\Lambda} \,.
\ee
where
the gauge parameter $\ket{\Lambda}$  for the
linearised gauge transformations is  a ghost number one state of the form:
\be
\label{gt-param}
\ket{\Lambda}  = {1\over g_s }
\int {d^D k \over (2\pi)^D}  ~ 
 {i\over \sqrt{2 }} \,
\lambda_\mu ( k) \, \bigl(  \alpha_{-1}^\mu  c_1
- \bar\alpha_{-1}^\mu  \bar c_1
 \bigr) \ket{k} \,.
\ee   
Expanding \refb{gtfa} 
gives  the following gauge transformations of the component fields:
\be
\label{collgt} 
\begin{split}
\delta h_{\mu\nu} &=   (\p_\mu
\lambda_\nu  +\p_\nu \lambda_\mu) \,, \\[1.0ex]
\delta f_\mu &= -\tfrac{1}{2} \,  \Box \,\lambda_\mu  \,,\\[1.0ex]
\delta e &=  -\tfrac{1}{2} \, \p \cdot\lambda \,. 
\end{split}
\ee
This can be confirmed to be a symmetry of the action \refb{quad-action}. 

We can now  eliminate
the auxiliary fields $f_\mu$
using their equations of motion: 
\be
\label{f-elim-quad}
f_\mu = -\tfrac{1}{2}\, 
 \bigl(  \p^\nu h_{\mu\nu} + 2 \p_\mu e\bigr) \,.
\ee
The result is the following quadratic action
\be
\label{eorkjr1}
\,S^{(2)} = {1\over 8  g_s^2} \int d^D x \,  ~L [ h, e] \,,  
 \ee
 where
 \be
\label{eorkjr}
L[\,h,e]
=   \tfrac{1}{4}   \, h^{\mu\nu} \Box\,   
h_{\mu\nu} 
+ \tfrac{1}{2}  (\partial^\nu h_{\mu\nu})^2-2\, e\, \partial^\mu \partial^\nu \, h_{\mu\nu}  
- 4\, e \,  \Box  
\, e  \,.
\ee
The gauge transformations  are 
\be
\label{lamtrans}
\begin{split}
\delta h_{\mu\nu}  &= ~  (\partial_\nu \lambda_\mu  +\partial_\mu \lambda_\nu)   \,,\\[1.0ex]
\delta e~ & = -\tfrac{1}{2}\,  \partial\cdot \lambda  
 \,.
\end{split}
\ee

We now compare this with the standard 
low-energy effective string theory
action $S_{\rm{st}}$ for gravity  and dilaton fields: 
\be 
\label{standardaction}
S_{\rm{st}}=  {1\over 2 \kappa^2} \int d^Dx  \sqrt{-{\rm g}}\, 
e^{-2\phi}  
 \Bigl[  R + 4 (\partial \phi)^2
\Bigr]\,.
\ee
Introducing fields $h_{\mu\nu},\phi$ via 
\be  \label{efieldexpand} 
{\rm g}_{\mu\nu} = \eta_{\mu\nu} + h_{\mu\nu},  
\qquad 
\phi = e + \tfrac{1}{4} \eta^{\mu\nu}h_{\mu\nu},
\ee 
and expanding the action to quadratic order in fluctuations, we get 
\be
\label{norm-action}
S^{(2)}_{\rm{st}} = {1\over 2 \kappa^2}\int d^Dx  \  L[\,h,e]\,,
\ee
the exact same result we had for the string field quadratic action, provided we make the
identification
\be   \label{ekappags}
\kappa = 2\, g_s \, .  
\ee
Eq.~\refb{efieldexpand} 
relates the string field fluctuations to the fluctuations
of the metric and dilaton.

Now, switching on a deformation \refb{e1pp4} corresponds to switching on 
$h_{\mu\nu}$
with $e=0$. This gives, from \refb{efieldexpand}
\be  \label{efieldexpandnew} 
{\rm g}_{\mu\nu} = \eta_{\mu\nu} + h_{\mu\nu},  
\qquad 
\phi =  \tfrac{1}{4} \eta^{\mu\nu}h_{\mu\nu}\, .
\ee 
This is the result used in \refb{ephiexp}.

\sectiono{Computation of D$p$-brane tension in bosonic string theory} \label{sb}

In this appendix we shall describe how the  equality of \refb{eTtwo} and \refb{eTthree}
can be used to compute the D$p$-brane tension in twenty six dimensional 
bosonic string theory.  
This is the open-closed SFT version of the classic computation 
given in~\cite{Polchinski:1998rq},
where Polchinski considered two separate D$p$-branes and compared open and closed 
channel versions of the annulus computation representing the interaction of the branes
due to exchange of strings.  Our computation is slightly more 
general, in that can be applied to space-filling D-branes which cannot be separated.  It
is also a bit simpler in the closed string channel, as one does not have to look at 
graviton exchanges; the tachyon exchange suffices.

From Ref.~\cite{Polchinski:1998rq} 
the annulus amplitude for
12 and 21 open strings on a pair of D-branes is  
\be
{\cal A}_{\rm ann}^P=   V_{p+1} \int_0^\infty {dt_p\over t_p}  (8\pi^2 
t_p)^{-(p+1)/2}  \eta(it_p)^{-24} \, .
\ee
We have dropped an $i$ from the result of \cite{Polchinski:1998rq} since 
we are working with an Euclidean path integral.
To pass to our case where  
we do not have separate  branes and thus 
no 12 and 21 strings, we have to make  
the replacement ${dt_p\over t_p} \to {dt_p\over 2t_p}$.  
Moreover $2\pi t_p = t$ for
us (see~\refb{eTthree}),
so we get
\be
{\cal A}_{\rm ann}=   
V_{p+1} \int_0^\infty {dt\over 2t}  (4\pi 
t)^{-(p+1)/2}  \Bigl[\eta\Bigl ({it\over 2\pi} \Bigr)\Bigr] ^{-24} \, .
\ee
Now use $t = 2\pi^2/s$  from \refb{eTfour} to rewrite  
\be
{\cal A}_{\rm ann}=  
V_{p+1} \int_0^\infty {ds\over 2s}  \Bigl({8\pi^3 
\over s} \Bigr)^{-(p+1)/2}  \Bigl[\eta\Bigl ({i\pi\over s} \Bigr)\Bigr] ^{-24} \, .
\ee
Since $\eta (iu) = u^{-1/2} \, \eta (i/u)$  we have
\be
\Bigl[\eta\Bigl ({i\pi\over s} \Bigr)\Bigr] ^{-24}  =  \Bigl( {\pi\over s} \Bigr)^{12} 
\Bigl[\eta\Bigl ({is\over \pi} \Bigr)\Bigr] ^{-24}  \simeq 
\Bigl( {\pi\over s} \Bigr)^{12}  \Bigl( e^{2s}  + 24 + \cdots\Bigr) \, ,
\ee
where on the right hand side we have expanded the integrand for large $s$. 
Therefore we have
\be
{\cal A}_{\rm ann}=  
V_{p+1} \int_0^\infty {ds\over 2s}  \Bigl({8\pi^3  
\over s} \Bigr)^{-(p+1)/2}  \Bigl( {\pi\over s} \Bigr)^{12}  \Bigl( e^{2s}  + 24 + \cdots\Bigr) \, , 
\ee
which simplifies to
\be\label{eaB2}
{\cal A}_{\rm ann}=  {  
V_{p+1} \over (8\pi^2 
)^{(p+1)/2}} \int_0^\infty {ds\over 2s}   \Bigl( {\pi\over s} \Bigr)^{(23-p)/2} 
 \Bigl( e^{2s}  + 24 + \cdots\Bigr) \,. 
\ee

Let us now evaluate the closed string version~\refb{eTtwo} of the annulus
partition function 
by inserting a complete set of states 
between $c_0^-$ and $b_0^+$ on the right hand side. This takes the form
\be
{\cal A}_{\rm ann}=-\tfrac{1}{2}  \, \eta_c \,  \int_0^\infty ds\,  \langle B| c_0^-\int {d^{26}k\over (2\pi)^{26}} \left[- c_1 \bar c_1|k\rangle \langle -k| c_{-1} \bar c_{-1} 
c_0 \bar c_0 +\cdots \right]b_0^+ \, e^{-s L_0^+} |B\rangle\, ,
\ee
where $\cdots$ denote contribution from states of higher $L_0^+$ eigenvalues
and the minus sign inside the square bracket is a reflection of the normalization condition
\refb{sl2c_overlap}. Using the fact that $c_0^+|B\rangle=0$,  and that 
$L_0^+ c_1\bar c_1 |k\rangle = -2 + k^2/2$, this expression can be
manipulated to
\be\label{eBk}
{\cal A}_{\rm ann}=\eta_c \,  \int ds\, \int {d^{26}k\over (2\pi)^{26}} \, e^{ - {1\over 2} s k^2 }\, 
\left[e^{2s} \langle B|  c_0^- c_1 \bar c_1|k\rangle \langle B|  c_0^- c_1 \bar c_1|-k\rangle
+\cdots\right] \, .
\ee
To proceed further, let us decompose $k$ as $(k_\parallel, k_\perp)$ where $k_\parallel$
is the component of the $p+1$ dimensional momentum along the D-brane world-volume
and $k_\perp$ is the component of the $(25-p)$ dimensional momentum transverse to the
D-brane world-volume. Then the matrix element $\langle B|  c_0^- c_1 \bar c_1|k\rangle$ is
independent of $k_\perp$ and we get,
\be
\int {d^{25-p}k_\perp\over (2\pi)^{25-p}} \, e^{-{1\over 2} s k_\perp^2}
= \left( {1\over2\pi s}\right)^{(25-p)/2}\, .
\ee
On the other hand, the $k_\parallel$ dependence of $\langle B|  c_0^- c_1 \bar c_1|k\rangle$
gives $(2\pi)^{p+1}\delta^{(p+1)}(k_\parallel)$. Therefore, after taking into account the
two such matrix elements in \refb{eBk}, the $k_\parallel$ integral gives
$(2\pi)^{(p+1)}\delta^{(p+1)}(k_\parallel=0)=V_{p+1}$.
Finally, 
the ghost part of the correlator $\langle B|
c_0^- c_1 \bar c_1|0\rangle$ 
can be represented as the correlation function
of $\langle \p c c \bar c(0)\rangle_{\rm disk}$. Using the map $z = i {1-iw\over 1+iw}$ from
the disk coordinate $w$ to the upper half plane coordinate $z$, we get
\ben
&&
\hskip-35pt
\langle\p c \, c \bar c (w=0)\rangle_{\rm disk} = {dw\over dz}\Bigl|_{w=0}\, 
{d\bar w \over d\bar z}\Bigl|_{w=0} \, 
\langle \p c \, c \bar c (z=i)\rangle_{\rm UHP}
\nonumber \\[1.2ex]
&=& \tfrac{1}{4} \langle \p c \, c(z=i) c(z=-i)\rangle_{\rm UHP} = K\, ,
\een
where we have used \refb{eopencorr} after stripping off the momentum dependent 
part that has already been accounted for. Using these results in \refb{eBk} we get,
\be\label{eaB1}
{\cal A}_{\rm ann} =
\eta_c\, K^2\, V_{p+1} \int_0^\infty ds \, \left( {1\over2\pi s}\right)^{(25-p)/2} \, [e^{2s}
+\cdots]\, . 
\ee
Comparing this determination of the annulus amplitude with that coming from the
open string picture in~\refb{eaB2} we finally fix the value of $K$:  
\be
\AL  K^2  =    {\pi\over 2^{12}} (4\pi^2)^{11-p} \,.
\ee
On the other hand, squaring the  relation \refb{eKTrelx} between $K$ and $\TT$, one has
\be   
\ \ \   \TT^2 =  {4\over g_s^2} \,  \AL \, K^2   \,.
\ee
Using our determination of $\AL K^2$ we then get   
\be
\TT^2 
=    {\pi\over 2^{10} g_s^2} (4\pi^2)^{11-p}  =   {\pi\over 256} { (4\pi^2)^{11-p} \over (2g_s)^2}\,.
\ee
This result for the D$p$-brane tension 
agrees with the result of~\cite{Polchinski:1998rq} (eqn.(8.7.26)) after using $\kappa=2\, g_s$ 
from~\refb{ekappags} and recalling that we work with $\alpha'=1$.


\end{document}